\documentclass[11pt,envcountsame]{llncs}
\usepackage{chngcntr}

\usepackage{url}
\usepackage[final]{graphicx}
\usepackage{amssymb}
\usepackage{amsmath}
\usepackage{algorithmic}
\usepackage[linesnumbered,ruled]{algorithm2e}
\usepackage{appendix}
\usepackage{hyperref}
\usepackage{bm}
\usepackage{cite}
\usepackage[papersize={8.5in,11in},top=1.2in,left=1.1in,right=1.1in,bottom=1.2in]{geometry}
\usepackage{xfrac}
\usepackage{enumitem}
\newenvironment{lessspaceitemize*}%
  {\begin{itemize}%
  \vspace{-1mm}
    \setlength{\itemsep}{0pt}%
    \setlength{\parskip}{0pt}}%
    {\vspace{-1mm} \end{itemize}}

\newenvironment{lessspaceenum*}%
  {\begin{enumerate}%
  \vspace{-1mm}
    \setlength{\itemsep}{0pt}%
    \setlength{\parskip}{0pt}}%
  {\end{enumerate}}

\counterwithout{figure}{section}

\newtheorem{clm}[theorem]{Claim}
\newenvironment{mainproof}{\paragraph{Proof:}}{\hfill $\blacksquare$\vskip 2pt}
\newenvironment{subproof}{\paragraph{Proof:}}{\hfill \qed\vskip 2pt}

\newenvironment{thm_app}[1]{\noindent\textbf{Theorem~\ref{#1}.}}{\par\addvspace{\baselineskip}}
\newenvironment{lem_app}[1]{\noindent\textbf{Lemma~\ref{#1}.}}{\par\addvspace{\baselineskip}}
\newenvironment{clm_app}[1]{\noindent\textbf{Claim~\ref{#1}.}}{\par\addvspace{\baselineskip}}
\newenvironment{prop_app}[1]{\noindent\textbf{Proposition~\ref{#1}.}}{\par\addvspace{\baselineskip}}

\spnewtheorem*{inf_thm}{(Informal Theorem)}{\bfseries}{\itshape}

\title{Computing Stable Coalitions:\\Approximation Algorithms for Reward Sharing}
\author{Elliot Anshelevich \and Shreyas Sekar}
\institute{Rensselaer Polytechnic Institute, Troy, NY}

\begin{document}


\maketitle

\begin{abstract}
Consider a setting where selfish agents are to be assigned to coalitions or projects from a set $\mathcal{P}$. Each project $k\in \mathcal{P}$ is characterized by a valuation function; $v_k(S)$ is the value generated by a set $S$ of agents working on project $k$. We study the following classic problem in this setting: ``how should the agents divide the value that they collectively create?". One traditional approach in cooperative game theory is to study \emph{core stability} with the implicit assumption that there are infinite copies of one project, and agents can partition themselves into any number of coalitions. In contrast, we consider a model with a finite number of non-identical projects; this makes computing both high-welfare solutions and core payments highly non-trivial.

The main contribution of this paper is a black-box mechanism that reduces the problem of computing a near-optimal core stable solution to the purely algorithmic problem of welfare maximization; we apply this to compute an approximately core stable solution that extracts one-fourth of the optimal social welfare for the class of subadditive valuations. We also show much stronger results for several popular sub-classes: anonymous, fractionally subadditive, and submodular valuations, as well as provide new approximation algorithms for welfare maximization with anonymous functions. Finally, we establish a connection between our setting
and the well-studied simultaneous auctions with item bidding; we adapt our results to compute approximate pure Nash equilibria for these auctions.
\end{abstract}


\setcounter{page}{1}

\section{Introduction}
``How should a central agency incentivize agents to create high value, and then distribute this value among them in a fair manner?" -- this question forms the central theme of this paper. Formally, we model a set of selfish agents in a combinatorial setting consisting of a set $\mathcal{P}$ of projects. Each project $k$ is characterized by a valuation function; $v_k(S)$ specifies the welfare generated by a set $S$ of agents working on project $k$. The problem that we study is the following: compute an assignment of agents to projects to maximize social welfare, and provide \emph{rewards} or \emph{payments} to each agent so that no group of agents deviate from the centrally prescribed solution.

For example, consider a firm dividing its employees into teams to tackle different projects. If these employees are not provided sufficient remuneration, then some group could break off, and form their own startup to tackle a niche task. Alternatively, one could imagine a funding agency  incentivizing researchers to tackle specific problems. More generally, a designer's goal in such a setting is to delicately balance the twin objectives of \emph{optimality} and \emph{stability}: forming a high-quality solution while making sure this solution is stable. A common requirement that binds the two objectives together is \emph{budget-balancedness}: the payments provided to the agents must add up to the total value of the given solution. \\

\noindent\textbf{Cooperative Coalition Formation} The question of how a group of agents should divide the value they generate has inspired an extensive body of research spanning many fields~\cite{bejanG09,hoefer13strategic,lewenbergBSZR15,myerson1977graphs,saad2009coalitional}. The notion of a `fair division' is perhaps best captured by the \textbf{Core}: a set of payments so that no group of agents would be better off forming a coalition by themselves. Although the Core is well understood, implicit in the papers that study this notion is the underlying belief that there are infinite copies of one single project~\cite{bachrachPR13,chalkiadakisEMPJ10}, which is often not realistic. For example, a tacit assumption is that if the payments provided are `not enough', then every agent $i$ can break off, and simultaneously generate a value of $v(i)$ by working alone; such a solution does not make sense when the number of projects or possible coalitions is limited. Indeed, models featuring selfish agents choosing from a finite set of \emph{distinct} strategies are the norm in many real-life phenomena: social or technological coordination~\cite{anshelevichS14,augustineCEFGS15}, opinion formation~\cite{chierichettiKO13,feldmanF15}, and party affiliation~\cite{balcanBM09,bhalgatCK10} to name a few.

The fundamental premise of this paper is that many coalition formation settings feature multiple non-identical projects, each with its own (subadditive) valuation $v_k(S)$. Although our model allows for duplicate projects, the inherently combinatorial nature of our problem makes it significantly different from the classic problem with infinite copies of a single project. For example, in the classic setting with a single valuation $v(S)$, the welfare maximization problem is often trivial (complete partition when $v$ is subadditive), and the stabilizing core payments are exactly the dual variables to the allocation LP~\cite{vetta2015coalition}. This is not the case in our setting where even the welfare maximization problem is NP-Hard, and known approximation algorithms for this problem use LP-rounding mechanisms, which are hard to reconcile with stability. Given this, our main contribution is a poly-time approximation algorithm that achieves stability without sacrificing too much welfare.


\subsection{The Core}
\label{sec_thecore}
Given an instance $(\mathcal{N}, \mathcal{P}, (v_k)_{k \in \mathcal{P}})$ with $N$ agents ($\mathcal{N}$) and $m$ projects, a solution is an allocation $S =(S_1, \ldots, S_m)$ of agents to projects along with a vector of payments $(\bar{p})_{i \in \mathcal{N}}$. With unlimited copies of a project, {\em core stability} refers to the inability of any set of agents to form a group on their own and obtain more value than the payments they receive.  The stability requirement that we consider is a natural extension of core stability to settings with a finite number of fixed projects. That is, when a set $T$ of agents deviate to project $k$, they cannot displace the agents already working on that project $(S_k)$. Therefore, the payments of the newly deviated agents (along with the payments of everyone else on that project) must come from the total value generated, $v_k(S_k \cup T)$. One could also take the Myersonian view~\cite{myerson1977graphs} that `communication is required for negotiation' and imagine that all the agents choosing project $k$ $(S_k \cup T)$ together collaborate to improve their payments. Formally, we define a solution $(S,\vec{\bar{p}})$ to be {\em core stable} if the following two conditions are satisfied,

\begin{description}
\item[(Stability)] No set of agents can deviate to a project and obtain more total value for everyone in that project than their payments, i.e., for every set of agents $T$ and project $k$, $\sum_{i \in T\cup S_k}\bar{p}_i \geq v_k(T\cup S_k)$.

\item[(Budget-Balance)] The total payments sum up to the social welfare (i.e., total value) of the solution: $\sum_{i \in \mathcal{N}}\bar{p}_i = \sum_{k \in \mathcal{P}}v_k(S_k)$.
\end{description}

\noindent Observe that Stability for $T=\emptyset$ together with budget-balancedness imply that the value created from a project will go to the agents on that project only. Finally, we consider a full-information setting as it is reasonable to expect the central authority to be capable of predicting the value generated when agents work on a project.

\vskip 3pt\noindent \textbf{(Example 1)} We begin our work with an impossibility result: even for simple instances with two projects and four agents, a core stable solution need not exist. Consider $\mathcal{P} = \left\lbrace 1,2 \right\rbrace$, and define $v_1(\mathcal{N}) = 4$ and $v_1(S) = 2$ otherwise; $v_2(S) = 1+\epsilon$ for all $S \subseteq N$. If all agents are assigned to project $1$, then in a budget-balanced solution at least one agent has to have a payment of at most $1$; such an agent would deviate to project $2$. Instead, if some agents are assigned to project $2$, then it is not hard to see that they can deviate to project $1$ and the total utility goes from $3+\epsilon$ to $4$.\\

\noindent\textbf{Approximating the Core} Our goal is to compute solutions that guarantee a high degree of stability. Motivated by this, we view core stability under the lens of approximation. Specifically, as is standard in cost-sharing literature~\cite{immorlicaMM08,roughgardenS09}, we consider relaxing one of the two requirements for core stability while retaining the other one. First, suppose that we generalize the Stability criterion as follows:

\begin{quote}
\textbf{($\alpha$-Stability)} For every set of agents $T$ and every project $k$, $v_k(S_k \cup T) \leq \alpha \sum_{i \in S_k \cup T}\bar{p}_i.$
\end{quote}

\noindent $\alpha$-stability captures the notion of a `switching cost' and is analogous to an \emph{Approximate Equilibrium}; in our example, one can imagine that employees do not wish to quit the firm unless the rewards are at least a factor $\alpha$ larger. In the identical projects literature, the solution having the smallest value of $\alpha$ is known as the Multiplicative Least-Core~\cite{vetta2015coalition}. Next, suppose that we only relax the budget-balance constraint,

\begin{quote}
\textbf{($\beta$-Budget Balance)} The payments are at most a factor $\beta$ larger than the welfare of the solution.
\end{quote}

\noindent This generalization offers a natural interpretation: the central authority can subsidize the agents to ensure high welfare, as is often needed in other settings such as public projects or academic funding~\cite{bejanG09}. 
In the literature, this parameter $\beta$ has been referred to as the \emph{Cost of Stability}~\cite{bachrach09cost,meirBR10}.

We do not argue which of these two relaxations is the more natural one: clearly that depends on the setting. Fortunately, it is not difficult to see that these two notions of approximation are equivalent. In other words, every approximately core stable solution with $\alpha$-stability can be transformed into a solution with $\alpha$-budget balancedness by scaling the payments of every player by a factor $\alpha$. Therefore, in the rest of this paper, we will use the term {\bf $\alpha$-core stable} without loss of generality to refer to either of the two relaxations. All our results can be interpreted either as forming fully budget-balanced payments which are $\alpha$-stable, or equivalently as fully stable payments which are $\alpha$-budget balanced. Finally, the problem that we tackle in this paper can be summarized as follows:

\begin{quote} \textbf{(Problem Statement)} Given an instance with subadditive valuation functions, compute an $\alpha$-core stable solution $(S, (\bar{p})_{i \in \mathcal{N}})$ having as small a value of $\alpha$ as possible, that approximately maximizes social welfare.
\end{quote}

\subsection{Our Contributions}


The problem that we face is one of bi-criteria approximation: to simultaneously optimize both social welfare and the stability factor $\alpha$ ($\alpha=1$ refers to a core stable solution). For the rest of this paper, we will use the notation $(\alpha, c)$-Core stable solution to denote an $\alpha$-Core solution that is also a $c$-Approximation to the optimum welfare. The bounds that we derive are quite strong: we are able to approximate both $\alpha$ and $c$ simultaneously to be close to the individually best-possible lower bounds. In a purely algorithmic sense, our problem can be viewed as one of designing approximation algorithms that require the additional property of \emph{stabilizability}. 

\noindent \textbf{Main Result} Our main result is the following black-box reduction that reduces the problem of finding an approximately core stable solution to the purely algorithmic problem of welfare maximization,

\begin{inf_thm}
For any instance where the projects have subadditive valuations, any LP-based $\alpha$-approximation to the optimum social welfare can be transformed in poly-time to a $(2\alpha, 2\alpha)$-core stable solution.
\end{inf_thm}

The strength of this result lies in its versatility: our algorithm can stabilize \emph{any input allocation} at the cost of half the welfare. The class of subadditive valuations is extremely general, and includes many well-studied special classes all of which use LP-based algorithms for welfare maximization; one can simply plug-in the value of $\alpha$ for the corresponding class to derive an approximately core stable solution. In particular, for general subadditive valuations, one can use the $2$-approximation algorithm of Feige~\cite{feige09subadd} and obtain a $(4,4)$-Core. As is standard in the literature \cite{dobzinskiNS10}, we assume that our subadditive functions are specified in terms of a demand oracle (see Section~\ref{sec:prelim} for more details). However, even in the absence of a demand oracle, one can obtain a poly-time reduction as long as we are provided an allocation and the optimum dual prices as input.

For various sub-classes of subadditive valuations, we obtain stronger results by exploiting special structural properties of those functions. These results are summarized in Table~\ref{table_results}. The classes that we study are extremely common and have been the subject of widespread interest in many different domains.

\begin{table}[htb]
\centering
\caption{Our results for different classes of complement-free valuations where $\text{Submodular} \subset \text{XoS} \subset \text{Subadditive}$, and $\text{Anonymous} \subset \text{Subadditive}$. The results are mentioned in comparison to known computational barriers for welfare maximization for the same classes, i.e., lower bounds on $c$.}
\begin{tabular}{|l|c | c|}
\hline
\textbf{Valuation Function Class} & \textbf{Our Results:} $(\alpha,c)$-Core & \textbf{Lower Bound for} $c$ \\
\hline
Subadditive & $(4,4)$ & $2$ \cite{feige09subadd}\\
\hline
Anonymous Subadditive & $(2,2)$ & $2$ \cite{feige09subadd}\\
\hline
Fractionally Subadditive (XoS) & $(1+\epsilon, \frac{e}{e-1})$ & $\frac{e}{e-1}$ \cite{dobzinskiNS10}\\
\hline
Submodular & $(1+\epsilon, \frac{e}{e-1})$ and $(1,2)$ & $\frac{e}{e-1}$ \cite{vondrak08}\\
\hline
\end{tabular}
\label{table_results}
\end{table}

\noindent\textbf{Lower Bounds.} All of our results are `almost tight' with respect to the theoretical lower bounds for both welfare maximization and stability. Even with anonymous functions, a $(2-\epsilon)$ core may not exist; thus our $(2,2)$-approximation for this class is tight. For general subadditive functions, one cannot compute better than a $2$-approximation to the optimum welfare efficiently, and so our $(4,4)$ result has only a gap of $2$ in both criteria. Finally, for XoS and Submodular functions, we get almost stable solutions ($(1+\epsilon)$-Core) that match the lower bounds for welfare maximization.

\noindent\textbf{A Fast Algorithm for Anonymous Subadditive Functions} We devise a greedy $2$-approximation algorithm for anonymous functions that may be of independent algorithmic interest. The only known $2$-approximation algorithm even for this special class is the rather complex LP rounding mechanism for general subadditive functions. In contrast, we provide an intuitive greedy algorithm that obtains the same factor, and use the structural properties of our algorithm to prove improved bi-criteria bounds ($(2,2)$ as opposed to $(4,4)$). \\

\noindent\textbf{Ties to Combinatorial Auctions with Item Bidding} We conclude by pointing out a close relationship between our setting and
\emph{simultaneous auctions} where buyers bid on each item separately~\cite{bhawalkarR11,christodoulouKS08, dobzinskiFK15}. Consider `flipping' an instance of our problem to obtain the following combinatorial auction: every project $k \in \mathcal{P}$ is a buyer with valuation $v_k$, and every $i \in \mathcal{N}$ is an item in the market. We prove an equivalence between Core stable solutions in our setting and Pure Nash equilibrium for the corresponding flipped simultaneous second price auction. Adapting our lower bounds to the auction setting, we make a case for Approximate Nash Equilibrium by constructing instances where every exact Nash equilibrium requires buyers to \emph{overbid} by a factor of $O(\sqrt{N})$, when they have anonymous subadditive valuations. Finally, we apply our earlier algorithms to efficiently compute approximate equilibria with small over-bidding for two settings, namely, $(i)$ a $\frac{1}{2}$-optimal, \emph{$2$-approximate equilibrium when buyers have anonymous subadditive valuations}, and $(ii)$ a $(1-\frac{1}{e})$-optimal, \emph{$1+\epsilon$-approximate equilibrium for submodular buyers}.

\subsection{Related Work}
The core has formed the basis for a staggering body of research in a myriad of domains, and one cannot hope to do justice to this vast literature. Therefore, we only review the work most pertinent to our model. The non-existence of the core in many important settings has prompted researchers to devise several natural relaxations~\cite{aumann1964bargaining,bachrach09cost,shapley1966quasi,schmeidler1969nucleolus}: of these, the Cost of Stability~\cite{bachrach09cost,bejanG09,meirBR10}, and the Multiplicative Least-Core~\cite{vetta2015coalition,schulzU13,shapley1966quasi} are the solution concepts that are directly analogous to our notion of an $\alpha$-core. That said, there are a few overarching differences between our model, and almost all of the papers studying the core and its relatives; $(i)$ \textbf{Duplicate Projects:} In the classic setting, it is assumed that there are infinite copies of one identical project so that
different subsets of agents (say $S_1$, $S_2$) working independently on the same project can each generate their full value $(v(S_1) + v(S_2))$, and $(ii)$ \textbf{Superaddivity:} In order to stabilize the grand coalition, most papers assume that the valuation is superadditive, which inherently favors cooperation. On the contrary, our setting models multiple dissimilar projects where each project is a fixed resource with a subadditive valuation.

Although cooperative games traditionally do not involve any optimization, a number of papers have studied well-motivated games where the valuation or cost function ($c(S)$) is derived from an underlying combinatorial optimization problem~\cite{deng1999algorithmic,goemans2004cooperative,hoefer13strategic,markakis2005core}. For example, in the vertex cover game~\cite{deng1999algorithmic,fangKZ08} where each edge is an agent, $c(S)$ is the size of the minimum cover for the edges in $S$. Such settings are fundamentally different from ours because the hardness arises from the fact that the value of the cost function cannot be obtained precisely. For many such problems, core payments can be computed almost directly using LP Duality~\cite{goemans2004cooperative,hoefer13strategic,markakis2005core}.

In the cooperative game theory literature, our setting is perhaps closest to the work studying coalitional structures where instead of forming the grand coalition, agents are allowed to arbitrarily partition themselves~\cite{bachrach09cost,grecoMPS11} or form overlapping coalitions~\cite{chalkiadakisEMPJ10}. This work has yielded some well-motivated extensions of the Core, albeit for settings with duplicate projects. Our work is similar in spirit to games where agents form coalitions to tackle specific tasks, e.g., threshold task games~\cite{chalkiadakisEMPJ10} or coalitional skill games~\cite{bachrachPR13}. In these games, there is still a single valuation function $v(S)$ which depends on the (set of) task(s) that the agents in $S$ can complete. Once again, the tacit assumption is that there are an infinite number of copies of each task.

Recently, there has been a lot of interest in designing cost-sharing mechanisms that satisfy strategy-proofness in settings where a service is to be provided to a group of agents who hold private values for the same~\cite{moulin2001strategyproof,devanurMV05}. In contrast,
we look at a full information game where the central agency can exactly estimate the output due to a set of agents working on a project. A powerful relationship between our work and the body of strategy-proof mechanisms was discovered by Moulin~\cite{moulin1999incremental} who showed that a natural class of `cross-monotonic cost sharing schemes' can be used to design mechanisms that are both core-stable (CS) and strategy-proof (SP). This has led to the design of beautiful SP+CS mechanisms for several combinatorially motivated problems with a single identical project or service~\cite{georgiou2013black,roughgardenS09}. Finally, we briefly touch upon the large body of literature in non-transferable utility games that (like us) study coalition formation with a finite number of asymmetric projects~\cite{augustineCEFGS15,branzeiL09,chierichettiKO13,feldmanLN12}. However, these papers use fixed reward-sharing schemes, and thus do not model the bargaining power of agents that is a key aspect of coalition formation.

\section{Model and Preliminaries}
\label{sec:prelim}
We consider a transferable-utility coalition formation game with a set $\mathcal{P}$ of $m$ projects and a set $\mathcal{N}$ of $N$ agents. Each project  $k \in \mathcal{P}$ is specified by a monotone non-decreasing valuation function $v_k : 2^\mathcal{N} \to \mathbb{R}^+ \cup \left\lbrace 0 \right\rbrace$. A solution consists of an allocation of agents to projects $S=(S_1,\ldots,S_m)$, and a payment scheme $(\bar{p})_{i \in \mathcal{N}}$ and is said to be $(\alpha,c)$-core stable for $\alpha \geq 1$, $c \geq 1$ if
\begin{itemize}
\item The payments are fully budget-balanced, and for every project $k$, and set $T$ of agents, $v_k(S_k \cup T) \leq \alpha \sum_{i \in S_k \cup T}\bar{p}_i$. An equivalent condition is that the payments are at most a factor $\alpha$ times the social welfare of the solution, and we have full stability, i.e., $v_k(S_k \cup T) \leq \sum_{i \in S_k \cup T}\bar{p}_i$.

\item The allocation $S$ is a $c$-approximation to the optimum allocation, i.e., the welfare of the solution $S$ is at least $\frac{1}{c}$ times the optimum welfare.
\end{itemize}

Throughout this paper, we will use $OPT$ to denote the welfare maximizing allocation as long as the instance is clear. Given an allocation $S = (S_1, \ldots, S_m)$, we use $SW(S)=\sum_{k=1}^m v_k(S_k)$ to denote the social welfare of this allocation, and $\zeta(S)$ to be the set of projects that are empty under $S$, i.e., $k \in \zeta(S)$ if $S_k = \emptyset$.

\subsection*{Comparison to Traditional Models} We digress briefly to highlight the key differences between our model as defined above and traditional utility-sharing settings found in the literature. Traditionally, a transferable-utility coalition formation game consists of a single valuation function $v(S)$. The objective there is to provide a vector of payments ($p_i$ to user $i$) in order to stabilize some desired solution $S=(S_1, \ldots, S_r)$\footnote{Usually, this is the grand coalition but it can also refer to other solutions, for example, the social welfare maximizing solution}, where the number of coalitions $r$ can be any positive integer. Here, core stability means that for any group of agents $T \subseteq \mathcal{N}$, $\sum_{i \in T}p_i \geq v(T)$. Notice from the above definition that (unlike our setting), the same core payments are applicable for every single solution $S$, i.e., the payments are completely independent of the solution formed. 

A stark contrast to our notion of a stable solution is the implicit assumption that there are an infinite number of copies of a single project (specified by $v(S)$) available for the agents to deviate to. For instance, a necessary condition for core stability is that $p_i \geq v(i)$ for every agent $i$; this implies that in theory, each of the $N$ agents could work independently on the same project and generate a total value of $\sum_{i}v(i)$ and not $v(N)$. As mentioned in the introduction, such assumptions do not always make sense, and it is reasonable to assume that the value generated depends on which project the agents deviate to, and how many other agents are currently working on that project or resource. Finally, in the traditional model, the minimum core payments (irrespective of the solution) can be obtained directly using the dual of the allocation LP. In contrast, this is not so in our setting due to the presence of slack variables (See Section~\ref{sec:main}).

\subsubsection*{Valuation Functions}
Our main focus in this paper will be on the class of monotone subadditive valuation functions. A valuation function $v$ is said to be subadditive if for any two sets $S, T \subseteq \mathcal{N}$, $v(S \cup T) \leq v(S) + v(T)$, and monotone if $v(S)\leq v(S\cup T)$. The class of subadditive valuations encompasses a number of popular and well-studied classes of valuations, but at the same time is significantly more general than all of these classes. It is worth noting that when there are an unlimited number of allowed groups, subadditive functions are almost trivial to deal with: both the maximum welfare solution and the stabilizing payments are easily computable. For our setting, however, computing OPT becomes NP-Hard, and a fully core-stable solution need not exist. Due to the importance and the natural interpretation of subadditive functions, we believe it is very desirable to understand utility sharing under such valuations; our paper presents the first known results on utility sharing for general subadditive functions. In addition, we are able to show stronger results for the following two sub-classes that are extremely common in the literature.

\begin{description}
\item [Submodular Valuations] For any two sets $S, T$ with $T \subseteq S$, and any agent $i$, $v(S \cup \left\lbrace i \right\rbrace) - v(S) \leq v(T \cup \left\lbrace i \right\rbrace) - v(T)$.

\item [Fractionally Subadditive (also called `XoS') Valuations] $\exists$ a set of additive functions $(a_1, \ldots, a_r)$ such that for any $T \subseteq \mathcal{N}$, $v(T) = \max_{j=1}^{r} a_j(T)$. These additive functions are referred to as {\em clauses.}
\end{description}

\noindent Recall that an additive function $a_j$ has a single value $a_j(i)$ for each $i \in \mathcal{N}$ so that for a set $T$ of agents, $a_j(T) = \sum_{i \in T}a_j(i)$. The reader is asked to refer to~\cite{dobzinskiNS10,feige09subadd,lehmannLN06,vondrak2007submodularity} for alternative definitions of the XoS class and an exposition on how both these classes arise naturally in many interesting applications.

\noindent\textbf{Anonymous Subadditive Functions} In project assignment settings in the literature modeling a number of interesting applications~\cite{kleinbergO11,meirBR10}, it is reasonable to assume that the value from a project depends only on the number of users working on that project. Mathematically, this idea is captured by anonymous functions: a valuation function is said to be anonymous if for any two subsets $S,T$ with $|S|=|T|$, we have $v(S) = v(T)$. One of our main contributions in this paper is a fast algorithm for the computation of Core stable solutions when the projects have anonymous subadditive functions. We remark here that anonymous subadditive functions form an interesting sub-class of subadditive functions that are quite different from submodular and XoS functions.

\subsubsection*{Demand Oracles.}
The standard approach in the literature while dealing with set functions (where the input representation is often exponential in size) is to assume the presence of an oracle that allows indirect access to the valuation by answering specific types of queries. In particular, when dealing with a subadditive function $v$, it is typical to assume that we are provided with a \emph{demand oracle} that when queried with a vector of payments $\vec{p}$, returns a set $T \subseteq \mathcal{N}$ that maximizes the quantity $v(T) - \sum_{i \in T}p_i$~\cite{dobzinskiNS10}. Demand oracles have natural economic interpretations, e.g., if $\vec{p}$ represents the vector of potential payments by a firm to its employees, then $v(T) - \sum_{i \in T}p_i$ denotes the assignment that maximizes the firm's revenue or surplus.

In this paper, we do not explicitly assume the presence of a demand oracle; our algorithmic constructions are quite robust in that they do not make any demand queries. However, any application of our black-box mechanism requires as input an allocation which approximates OPT, and the optimum dual prices, both of which cannot be computed without demand oracles. For example, it is well-known~\cite{dobzinskiNS10} that one cannot obtain any reasonable approximation algorithm for subadditive functions (better than $O(\sqrt{N})$) in the absence of demand queries. That said, for several interesting valuations, these oracles can be constructed efficiently. For example in the case of XoS functions, a demand oracle can be simulated in time polynomial in the number of input clauses. We conclude this discussion by reiterating that demand oracles are an extremely standard tool used in the literature to study combinatorial valuations; almost all of the papers~\cite{dobzinskiNS10,georgiou2013black,feige09subadd} studying Subadditive or XoS functions take the presence of a demand oracle for granted.

\subsection{Warm-up Result: $(1,2)$-Core for Submodular Valuations}
We begin with an easy result: an algorithm that computes a core stable solution when all projects have submodular valuations, and also retains half the optimum welfare. Although this result is not particularly challenging, it serves as a useful baseline to highlight the challenges involved in computing stable solutions for more general valuations. Later, we show that by sacrificing an $\epsilon$ amount of stability, one can compute for submodular functions, a solution with a much better social welfare $(\frac{e}{e-1}$-approximation to OPT).
\begin{clm}
\label{clm_submodular_warmup}
We can compute in poly-time a $(1,2)$-Core stable solution for any instance with submodular project valuations.
\end{clm}
The above claim also implies that for every instance with submodular project valuations, there exists a Core stable solution. In contrast, for subadditive valuations, even simple instances (Example 1) do not admit a Core stable solution.
\begin{mainproof}
The proof uses the popular greedy half-approximation algorithm for submodular welfare maximization due to~\cite{lehmannLN06}. Initialize the allocation $X$ to be empty. At every stage, add an agent $i$ to project $k$ so that the value $v_k(X_k \cup \left\lbrace i \right\rbrace) - v_k(X_k)$ is maximized. Set $i$'s final payment $\bar{p}_i$ to be exactly the above marginal value. Let the final allocation once the algorithm terminates be $S$, so $\sum_{i\in S_k}\bar{p}_i = v_k(S_k)$. Consider any group of agents $T$, and some project $k$: by the definition of the greedy algorithm, and by submodularity, it is not hard to see that $\forall i \in T$, $\bar{p}_i \geq v_k(S_k\cup \left\lbrace i \right\rbrace) - v_k(S_k)$. Therefore, we have that $\sum_{i \in T}\bar{p}_i \geq v_k(S_k \cup T) - v_k(S_k)$, and since the payments are clearly budget-balanced, the solution is core-stable.
\end{mainproof}

\noindent\textbf{Challenges and Techniques for Subadditive Valuations} At the heart of finding a Core allocation lies the problem of estimating `how much is an agent worth to a coalition'. Unfortunately, the idea used in Claim~\ref{clm_submodular_warmup} does not extend to more general valuations as the \emph{marginal value} is no longer representative of an agent's worth. One alternative approach is to use the dual variables to tackle this problem: for example, in the classic setting with duplicate projects, \textbf{every} solution $S$ along with the dual prices as payments yields an $\alpha$-budget balanced core. Therefore, the challenge there is to bound the factor $\alpha$ using the integrality gap. However, this is no longer true in our combinatorial setting as the payments are closely linked to the actual solution formed, and moreover, there is no clear way of dividing the dual variables due to the presence of \emph{slack} (see LP~\ref{lp_welfaremaximization}). 

\noindent \textbf{Our Approach.} We attempt to approximately resolve the question of finding each agent's worth by identifying (for each project) a set of ``heavy users" who contribute to half the project's value. We provide large payments to each heavy user based on her \emph{best outside option} which are determined using Greedy Matchings. Finally, the dual variables are used only as a `guide' to ensure that $\forall k \in \mathcal{P}$, the payment given to the users on that project is at least a good fraction of the value they generate.

\section{Computing Approximately Core Stable Solutions}
\label{sec:main}
In this section, we show our main algorithmic result, namely a black-box mechanism that reduces the problem of finding a core stable solution to the algorithmic problem of subadditive welfare maximization. We use this black-box in conjunction with the algorithm of Feige~\cite{feige09subadd} to obtain a $(4,4)$-Core stable solution, i.e., a $4$-approximate core that extracts one-fourth of the optimum welfare. Using somewhat different techniques, we form stronger bounds ($(2,2)$-Core) for the class of anonymous subadditive functions. Our results for the class of anonymous functions are tight: there are instances where no $(2-\epsilon, 2-\epsilon)$-core stable solution exists. This indicates that our result for general subadditive valuations is close to tight (up to a factor of two).

We begin by stating the following standard linear program relaxation for the problem of computing the welfare maximizing allocation. Although the primal LP contains an exponential number of variables, the dual LP can be solved using the Ellipsoid method where the demand oracle serves as a separation oracle~\cite{dobzinskiNS10}. The best-known approximation algorithms for many popular classes of valuations use LP-based rounding techniques; of particular interest to us is the $2$-approximation for Subadditive valuations~\cite{feige09subadd}, and $\frac{e}{e-1}$-approximation for XoS valuations~\cite{dobzinskiNS10}.

\begin{equation}
\begin{aligned}
\quad \text{max} & \quad \sum_{k=1}^M \sum_{S\subseteq \mathcal{N}} x_k(S)v_k(S)
& (D) \quad \text{min} & \quad \sum_{i=1}^N p_i + \sum_{k=1}^M z_k & \\
\text{s.t.} & \quad \sum_{k=1}^M \sum_{S \ni i}x_k(S) \leq 1 \quad \forall i \in \mathcal{N}
&\text{s.t.} & \quad \sum_{i \in S}p_i + z_k \geq v_k(S) \quad \forall S,k \\
& \quad \quad \sum_{S \subseteq \mathcal{N}}x_k(S) \leq 1, \quad \forall k \in \mathcal{P}
& \quad \quad p_i & \geq 0, \quad \forall i \in \mathcal{N}\\
& \quad \quad x_k(S) \geq 0, \quad \forall S, \forall k
& \quad \quad z_k & \geq 0, \quad \forall k \in \mathcal{P}
\end{aligned}
\label{lp_welfaremaximization}
\end{equation}

As long as the instance is clear from the context, we will use $(\vec{p^*}, \vec{z^*})$ to denote the optimum solution to the Dual LP, referring to $\vec{p^*}$ as the dual prices, and $\vec{z^*}$ as the slack.

\noindent\textbf{Main Result} We are now in a position to show the central result of this paper. The following black-box mechanism assumes as input an \emph{LP-based $\alpha$-approximate allocation}, i.e., an allocation whose social welfare is at most a factor $\alpha$ smaller than the value of the LP optimum for that instance. LP-based approximation factors are a staple requirement for black-box mechanisms that explicitly make use of the optimum LP solution~\cite{georgiou2013black}. Along these lines, we make the assumption that the optimum dual variables (for the given instance) are available to the algorithm along with an input allocation.

\begin{theorem}
\label{thm_mainsubadditive}
Given any $\alpha$-approximate solution to the LP optimum, we can construct a $(2\alpha, 2\alpha)$-Core Stable Solution in polynomial time as long as the projects have subadditive valuations.
\end{theorem}

For general subadditive functions, the only known poly-time constant-factor approximation is the rather intricate randomized LP rounding scheme proposed in~\cite{feige09subadd}. Using this $2$-approximation, we get the following corollary. 

\begin{corollary}
We can compute in poly-time a $(4,4)$-Core stable solution for any instance with subadditive projects.
\end{corollary}
\noindent We now prove Theorem~\ref{thm_mainsubadditive}. \begin{mainproof}
We provide an algorithm that takes as input an allocation $A=(A_1,\ldots A_m)$ that is an $\alpha$-approximation to the LP Optimum and returns a core stable solution $S=(S_1,\ldots, S_m)$ along with payments $(\bar{p})_{i \in N}$ whose welfare is at least half that of $A$, and such that the total payments are at most $2\alpha$ times the welfare of $S$. 

Recall that in a core stable solution $S$, it is necessary that for every project $k$ and set $T$ of agents, $v_k(S_k \cup T) \leq \sum_{i \in S_k \cup T}\bar{p}_i$. A naive approach is to consider whether the dual payments (price $p^*_i$ plus the slack $z^*_k$ divided equally among $A_k$) would suffice to enforce core stability to the solution $A$. Unfortunately, this naive strategy fails because the payments are not enough to prevent the deviation of agents to empty projects. To remedy this, we take the following approach: we implement a matching-based routine that allows us to identify the `light' and `heavy' users at each project so that when the light users deviate to the empty projects, there is not much welfare loss (and vice-versa for the heavy users). We assign the light users to these projects, and provide the heavy users with payments that depend on both `the best outside option' available to them and their contribution to social welfare in order to stabilize them. 

We begin by defining a simple \emph{Greedy Matching with Reserve Prices} procedure that will serve as a building block for our main algorithm. The procedure is straightforward so we state it in words here and formally define it in Appendix~\ref{app:mainsub}. 
\begin{quote}\textbf{Algorithm 2:} ``Begin with an input allocation and initial payments. During every iteration, assign an agent $i$ to a currently empty project $k$, as long as her current payment $p_i < v_k(i)$, and update her payment to $v_k(i)$. Terminate when $p_i \geq v_k(i)$ for each agent $i$ and empty project $k$."
\end{quote}


%

\noindent We begin our analysis of the above procedure with a simple observation: during the course of the algorithm, the payments of the agents are non-decreasing (in fact, in every iteration, the payment of at least one agent strictly increases). Specifically, we are interested in analyzing the solution returned by the algorithm when the input allocation is $A$, and the input payments are the naive dual payments discussed above. We first describe some notation and then prove some lemmas regarding the solution returned by the algorithm for this input. Recall that for any given solution $X$, $\zeta(X)$ denotes the set of empty projects under $X$.

We denote by $\vec{p^0}$ the marginal contributions given by the optimal dual payment plus the slack divided equally as per $A$, i.e., if agent $i \in A_k$, then $p^0_i = p^*_i + \frac{z^*_k}{|A_k|}$. Suppose we run the algorithm on the input $(A,\vec{p^0})$; let the corresponding output allocation be $B$, and the payments be $p^B$. Also define for every $k \in \mathcal{P} \setminus \zeta(A)$, $A_k^+=A_k\cap B_k$ to be the agents who remained on project $k$, $A_k^-=A_k\setminus A^+_k$ to be the agents who left project $k$, and $P_k$ to be the set of projects that the agents in $A^-_k$ switched to in allocation $B$. Note that all the projects in $P_k$ will only have one agent each in $B$ due to the definition of the algorithm. We now divide the non-empty projects in $A$ into two categories based on the welfare lost after running the algorithm. Specifically, consider any project $k$ in $\mathcal{P} \setminus \zeta(A)$. We refer to $k$ as a \textbf{good} project if the welfare in $B$ due to the agents originally in $A_k$ is at least half their original welfare, and refer to $k$ as a \textbf{bad} project otherwise. That is, $k$ is a \emph{good} project iff,
$$v_k(A^+_k) + \sum_{l \in P_k} v_l(B_l) \geq \frac{v_k(A_k)}{2}.$$

The following lemma which we prove in the Appendix establishes the crucial fact that although bad projects may result in heavy welfare losses, they surprisingly retain at least half the agents originally assigned to them under $A$. Later, we use this to infer that the agents who deviated from bad projects are `heavy' users who contribute significantly to the project's welfare.

\begin{lemma}
\label{lem_subadd_halfagentsgone}
For every bad project $k$, $|A^+_k| > |A^-_k|$, i.e., more than half the agents in $A_k$ still remain in project $k$.

\end{lemma}

Our next lemma relates the output payments $\vec{p^B}$ to the optimal dual variables.

\begin{lemma}
\label{lem_match_upperbound}
For every project $k$, and every agent $i$ who is allocated to $k$ in $B$, her payment under $\vec{p^B}$ is not larger than $p^*_i + \frac{z^*_k}{|B_k|}$.
\end{lemma}
\begin{subproof}
We prove the lemma in two cases. First consider any agent $i$ whose allocation remains the same (say project $k$) during the entire course of Algorithm~\ref{alg_greedy} for the input $(A,\vec{p^0})$. Clearly, this agent's final payment returned by the algorithm $p^B_i$ is exactly the same as her initial payment $p^0_i = p^*_i + \frac{z^*_k}{|A_k|}$. However, we know that $|A^+_k| \leq B_k$. Therefore, $p^B_i = p^*_i + \frac{z^*_k}{|A_k|} \leq p^*_i + \frac{z^*_k}{|B_k|}$.

Next, consider an agent $i \in B_k$ whose allocation changed at some point during the course of the algorithm. This means that $|B_k|=1$. Then, by definition, her final payment is exactly $v_k(i)$, which is not larger than $p^*_i + z^*_k$ by dual feasibility.
\end{subproof}

\textbf{Main Algorithm: Phase I} While the returned solution $B$ is indeed core stable, its welfare may be poor due to the presence of one or more \emph{bad} projects. Instead of using solution $B$, we use its structure as a guide for how to form a high-welfare solution. For {\em good} projects, we can put the agents in $A_k^+$ onto $k$ and the agents in $A_k^-$ onto $P_k$; since these are good projects this is guaranteed to get us half of the welfare $v_k(A_k)$, as desired. For {\em bad} projects, on the other hand, more than half of the welfare disappeared when we moved agents on $A^-_k$ away; due to sub-additivity this means that $v_k(A^-_k)\geq \frac{v_k(A_k)}{2}$. So instead we will assign agents in $A^-_k$ to project $k$ (which is the opposite of what happens in solution $B$), and put some agents from $A^+_k$ onto projects $P_k$. This is Phase I of our main algorithm, defined formally in Appendix~\ref{app:mainsub}. 

\noindent\textbf{Payments at the end of Phase I:} Suppose that the allocation at the end of the above procedure is $S'$; let us define the following payment vector $\vec{p'}$. For every good project $k$: for each agent $i$ assigned to $l \in  k \cup P_k$, her payment is $p'_i = p^*_i + \frac{z^*_l}{|S'_l|}.$ For every bad project $k$, define $D_k:= S'_k \setminus A^-_k$ to be the set of dummy agents belonging to that project. Each dummy agent receives exactly $p'_i=p^*_i$ as payment; every non-dummy agent assigned to a bad project receives $p'_i = p^B_i$ plus the left over slack from that project. For each bad project $k$, every agent $i$ assigned to some $l \in P_k$ receives a payment of $p'_i = p^*_i + z^*_l$.

We break the flow of our algorithm and show some properties satisfied by the solution returned by Phase I of our algorithm $(S', \vec{p'})$. Mainly we show that this solution is \emph{almost} core-stable and has desired welfare properties. In Phase II, we once again invoke our Greedy Matching Procedure to ensure core-stability. Recall that every \emph{bad} project contains at least one dummy agent; all the agents $N$ other than the dummy agents will be referred to as \emph{non-dummy} agents. 

\begin{lemma}
\label{lem_subadd_nondummy}
For every agent $i$ that does not belong to the set of dummy agents, her payment at the end of the first phase ($p'_i$) is at least her payment returned by the call to the Greedy Matching Procedure $p^B_i$.
\end{lemma}

Specifically, the above lemma implies that with respect to the non-dummy agents, our solution $(S',\vec{p'})$ retains the `nice' stability properties guaranteed by the greedy matching procedure.

\begin{corollary}
\label{corr_subadd_emptyproj}
For every empty project $k$ in $S'$ (i.e., $k \in \zeta(S')$), and every non-dummy agent $i$, her payment at the end of the first phase is at least her individual valuation for project $k$, i.e.,

$$p'_i \geq v_k(i).$$
\end{corollary}
Now that we have a lower bound on the payments returned by the first phase of our algorithm, we show a stricter lemma giving an exact handle on the payments.


\begin{lemma}
\label{lem_paymentfirstphase}
For every non-empty project $k \notin \zeta(S')$, the total payment to agents in $k$ at the end of Phase I is exactly $\sum_{i \in S'_k}p^*_i + z^*_k$.
\end{lemma}

Our final lemma shows that the total welfare at the end of the first phase is at least half the welfare of the original input allocation $A$.

\begin{lemma}
\label{lem_welfare}
For every \emph{good} project $k$, the welfare due to the agents in $k \cup P_k$ is at least half of $\frac{v_k(A_k)}{2}$. For every \emph{bad} project $k$, the welfare due to the non-dummy agents in $k$, i.e., $A^-_k$ is at least half of $\frac{v_k(A_k)}{2}$.
\end{lemma}
\begin{subproof}
The first half of the lemma is trivially true because of the definition of \emph{good} projects and the fact that the allocation of agents to the projects in $k \cup P_k$ is the same as the allocation returned by the call to Algorithm~\ref{alg_greedy}.

Moving on to \emph{bad} projects, we know that $S'_k \supseteq A^-_k$ and the agents in $A^-_k$ are exactly the non-dummy agents in project $k$. Therefore, we have
\begin{align*}
v_k(A^-_k) & \geq v_k(A_k) - v_k(A^+_k) \geq v_k(A_k) - \frac{v_k(A_k)}{2} & \quad \text{(By the definition of \emph{bad} projects)}.
\end{align*}
Observe that by virtue of this lemma, we can immediately obtain that the solution returned by the first phase of our algorithm has half the social welfare of the allocation $A$.
\end{subproof}

\subsection*{Main Algorithm - Phase II}
From the above lemmas, it is not hard to conclude that the solution $S'$ at the end of Phase I has good social welfare and is resilient against deviations to empty projects as long as we only consider non-dummy agents\footnote{We can also show that the solution is resilient against deviations to non-empty projects, although this is not needed at this time.}. In the second phase of our algorithm, we fix this issue by allowing dummy agents to deviate to empty projects using our Greedy Matching procedure and lower bounding their final payments using the dual variables. We formalize the algorithm for Phase II in the Appendix. Suppose that $S$ is the solution returned by the Greedy Matching Algorithm with input $(S',\vec{p'})$, and $\vec{\bar{p}}$ is the payment vector where all the agents who deviated from $S'$ receive as much as their dual variables, and the rest of the agents receive their payment under $\vec{p'}$. We now state some simple properties that compare the output of Phase II with its input, and formally prove them in Appendix~\ref{app:mainsub}.
\begin{clm}
\label{clm_phase2invariants}
The following properties are true:
\begin{enumerate}
\item The set of empty projects in $S$ is a subset of the set of empty projects in $S'$, i.e., $\zeta(S) \subseteq \zeta(S')$.
\item For all non-dummy agents, their strategies in $S'$ and $S$ coincide.
\item For every agent $i \in N$, her payment at the end of Phase II ($\bar{p}_i$) is at least her payment at the end of Phase I.

\end{enumerate}
\end{clm}

Our final lemma before showing the main theorem tells us that for every project, the total payment made to agents of that project coincides with the dual payments. The final payments to agents, therefore, are simply a redistribution of the dual payments. We defer its proof to the Appendix.
\begin{lemma}
\label{lem_paymentbounds}
For every non-empty project $k \notin \zeta(S)$, the total payments made to agents in $k$ is exactly $\sum_{i \in S_k}p^*_i + z^*_k$. Moreover, the payment made to any agent $i$ is at least her dual price $p^*_i$.
\end{lemma}

The rest of the theorem follows almost immediately. We begin by showing that the solution $(S, (\bar{p})_i)$ is \textbf{core stable}. Consider any project $k$, and a deviation by some set of agents $T$ to this project. We only have to show that the total payment made to the agents in $S_k \cup T$ is at least $v_k(S_k \cup T)$. We proceed in two cases.

%

First, suppose that $S_k = \emptyset$. Then, we have $v_k(S_k \cup T) \leq \sum_{i \in S_k \cup T} v_k(i) \leq \sum_{i \in S_k \cup T} \bar{p}_i$ from Subadditivity, and Claim~\ref{clm_phase2invariants} respectively. Claim~\ref{clm_phase2invariants}). Now suppose that $S_k \neq \emptyset$. Then, we have
\begin{align*}
\sum_{i \in S_k \cup T}\bar{p}_i = \sum_{i \in S_k} \bar{p}_i + \sum_{i \in T}\bar{p}_i & \geq \sum_{i \in S_k}p^*_i + z^*_k + \sum_{i \in T}p^*_i & \text{By Lemma~\ref{lem_paymentbounds}}\\
& \geq v_k(S_k \cup T) & \text{By dual feasibility}
\end{align*}

We now establish that the social welfare of our solution is at least half the social welfare of the original allocation $A$. Recall that every non-empty project in $A$ was classified as a \emph{good} or \emph{bad} project. For every \emph{good} project $k$ and its associated projects $P_k$, the fact that $v_k(S_k) + \sum_{l \in P_k}v_l(S_l) \geq v_k(A_k)/2$ follows from Lemma~\ref{lem_welfare} since $S_k = S'_k$, and $S_l = S'_l$. 

Consider any \emph{bad} project $k$. We know that for every non-dummy agent in $S'_k$, her strategy in $S$ is still project $k$. Therefore, the welfare due to any \emph{bad} project is at least the welfare due to the non-dummy agents in that project which by Lemma~\ref{lem_welfare} is at least half of $v_k(A_k)$. Finally, all that remains is to show that the total payments made in $(\bar{p})_i$ are at most a factor $2\alpha$ larger than the welfare of the solution.

From Lemma~\ref{lem_paymentbounds}, we know that the total payments made to agents at the end of Phase I is at most the value of the Dual Optimum of LP~\ref{lp_welfaremaximization}, which by Strong Duality is equal to the value of the Primal Optimum. However, we know that the welfare of $S$ is at least half the welfare of $A$, which by definition is at most a factor $\alpha$ away from the LP Optimum. This completes the proof.
\end{mainproof}

\subsection{Anonymous Functions}
Our other main result in this paper is a $(2,2)$-Core stable solution for the class of subadditive functions that are anonymous. Recall that for an anonymous valuation $v$, $v(T_1) = v(T_2)$ for any $|T_1|=|T_2|$. Such functions are frequently assumed in coalition formation and project assignment settings~\cite{kleinbergO11}. We begin with some existential lower bounds for approximating the core. From Example (1), we already know that the core may not exist even in simple instances. Extending this example, we show a much stronger 
set of results.

\begin{clm}
\label{clm_badexamplesubadd}
\textbf{(Lower Bounds)} There exist instances having only two projects with anonymous subadditive functions such that
\begin{enumerate}

\item For any  $\epsilon > 0$, no $(2-\epsilon, c_1)$-core stable solution exists for any value $c_1$.

\item For any  $\epsilon > 0$, no $(c_2, 2-\epsilon)$-core stable solution exists for any value $c_2$.
\end{enumerate}
\end{clm}
\emph{(Proof Sketch)} (Part 1) For ease of notation, we show that no $(2-\epsilon)$-budget-balanced core stable solution exists for a given $\epsilon > 0$. Consider an instance with $N$ buyers. The valuations for the two projects are $v_1(S) = \frac{N}{2} ~~ \forall S \subset \mathcal{N}$, and $v_1(\mathcal{N}) = N$; $v_2(S) =  2 ~~ \forall S \subseteq N$. Assume by contradiction that there is a $(2-\epsilon)$-core stable solution, then this cannot be achieved when all of the agents are assigned to project $1$ because they would each require a payment of $2$ to prevent them from deviating to project $2$. On the other hand, suppose that some agents are assigned to project $2$, then the social welfare of the solution is at most $\frac{N}{2} + 2$. If these agents cannot deviate to project $1$, then, the total payments would have to be at least $v_1(\mathcal{N}) = N$. For a sufficiently large $N$, we get that the budget-balance is $\frac{N}{N/2 + 2} > 2-\epsilon$. The example for Part 2 is provided in the Appendix. \hfill $\blacksquare$ \vskip3pt

We now describe an intuitive $2$-approximation algorithm (Algorithm~\ref{alg_greedyanon}) for maximizing welfare that may be of independent interest. To the best of our knowledge, the only previously known approach that achieves a $2$-approximation for anonymous subadditive functions is the LP-based rounding algorithm for general subadditive functions~\cite{feige09subadd}. Our result shows that for the special class of anonymous functions, the same approximation factor can be achieved by a much faster, greedy algorithm. In addition, our greedy algorithm also possesses other `nice structural properties' that may be of use in other settings such as mechanism design~\cite{lucierB10}.


\begin{algorithm}[htbp]
\caption{Greedy $2$-Approximation Algorithm for Anonymous Subadditive Functions}
\label{alg_greedyanon}
\algsetup{indent=2em}
\begin{algorithmic}[1]
\STATE Initialize the set of unallocated agents $U \gets \mathcal{N}$
\STATE Initialize the current allocation $S \gets (\emptyset, \ldots, \emptyset)$
\WHILE{$U \neq \emptyset$}
\STATE Find a set $T \subseteq U$ that maximizes the ratio $\displaystyle \frac{v_k(T | S_k)}{|T|}$ over all $k \in \mathcal{P}$
\STATE Assign the agents in $T$ to the project $k$ that maximizes the above ratio
\COMMENT{$S_k = S_k \cup T$ and $U = U \setminus T$.}
\STATE For all $i \in T$, set agent $i$'s marginal contribution $p_i = \frac{v_k(T | S_k)}{|T|}.$
\ENDWHILE
\STATE Return the final allocation $S$.
\end{algorithmic}
\end{algorithm}

\noindent{\bf Recall that the quantity $v(T | S)$ refers to $v(S \cup T) - v(S)$.}\vskip 2pt



Although Algorithm~\ref{alg_greedyanon} is only an approximation algorithm, the following theorem shows that we can utilize the greedy structure of the allocation and devise payments that ensure core stability. In particular, the solution that we use to construct a (yet to be proved) $(2,2)$-core is $(S, \vec{\bar{p}})$, where $S$ is the allocation returned by the algorithm, and $\bar{p}_i = 2p_i$ is the payment provided to agent $i$. We remark that the `marginal contributions' are defined only for the sake of convenience. They do not serve any other purpose. We make the following simple observation regarding the algorithm: the total social welfare of the solution $S$, $SW(S)$ is exactly equal to the sum of the marginal contributions $\sum_{i \in N}p_i$.

\begin{theorem}
\label{thm_anonsubadditive}
For any instance with anonymous subadditive projects, the allocation $S$ returned by Algorithm~\ref{alg_greedyanon} along with a payment of $\bar{p}_i = 2p_i$ for every $i \in \mathcal{N}$ constitutes a $(2,2$)-core stable solution.
\end{theorem}
\begin{mainproof}
We begin with some basic notation and simple lemmas highlighting the structural properties of our algorithm leading up to the main result. First, note that the total payments are exactly equal to the twice the aggregate marginal contribution, which in turn is equal to twice the social welfare. Therefore, our solution is indeed $2$-budget balanced. Now, let us divide the execution of the greedy algorithm into iterations from $1$ to $r$ such that in every iteration, the algorithm chooses a set of unallocated agents maximizing the marginal contribution (average increase in welfare). We define $A_j$ to be the set of agents assigned to some project during iteration $j \leq r$. Clearly, all the agents in $A_j$ are allocated to the same project, and have the exact same marginal contribution, and therefore payment. Let us use $p^{(j)}$ to refer to the marginal contribution of the agents in $A_j$.

Note that in order to characterize the \emph{state of the algorithm} during iteration $j$, it is enough if we express the set of agents assigned to each project, and the set of unallocated agents. Define $S^{(j)}_k$ to be the set of agents allocated to project $k$ at the beginning of iteration $j$ (before the agents in $A_j$ are assigned), and $U^{(j)}$ to be the set of unallocated agents during that instant. Suppose that the agents in $A_j$ are assigned to project $k$, then by definition the following equation must be true,

$$p^{(j)} = \frac{v_{k}(A_j | S^{(j)}_{k})}{|A_j|}.$$

Finally, given any $T \subseteq N$, we denote by $T^>$, the ordered set of elements in $T$ in the decreasing order of their payment. We begin with a simple property that links the prices to the welfare of every project.

\begin{proposition}
\label{prop_paymentwelfare}
In the final allocation $S$, the social welfare due to every project $k$ equals the total marginal contributions to the agents assigned to that project, i.e.,
$$\sum_{i \in S_k}p_i = v_k(S_k).$$
\end{proposition}
The proof follows directly from the definition of the algorithm. Now, we establish that as the algorithm proceeds, the marginal contributions of the agents cannot increase.

\begin{lemma}
\label{lem_ano_monoton}
For every $i,j$ with $i < j$, the marginal of the agents in $A_i$ is not smaller than the marginal of the agents in $A_j$, i.e., $p^{(1)} \geq p^{(2)} \geq \ldots \geq p^{(r)}$.
\end{lemma}
\emph{(Proof Sketch)} Since, the assignment of agents to any one project does not affect the marginal contribution (or average increase in welfare) in other projects, it suffices to prove the lemma for the case when $A_i$ and $A_j$ are assigned to the same project. The rest of the proof involves showing that if the lemma does not hold, then adding $A_i \cup A_j$ instead of $A_i$ in iteration $i$ would have lead to larger average welfare. The full proof is in the Appendix. \hfill $\blacksquare$

Recall that Proposition~\ref{prop_paymentwelfare} equates the marginal contributions to the welfare for every set $S_k$. The following lemma establishes a relationship between payments and welfare for subsets of $S_k$. Note that since the payments to the agents are exactly twice their marginal, we can use the payments and marginal contributions interchangeably. Once again, its proof is in Appendix~\ref{app:anon}.

\begin{lemma}
\label{lem_welfarepayupperb}
For every project $k$, and any given positive integer $t \leq |S_k|$, the total marginal contribution of the $t$ highest paid agents in $S_k$ is at least the value derived due to any set of $t$ agents, i.e., if $T$ denotes the set of $t$-highest paid agents in $S_k$, then

$$\sum_{i \in T}p_i \geq v_k(T).$$
\end{lemma}

We now move on to the most important component of our theorem, which we call the Doubling Lemma. This lemma will serve as the fundamental block required to prove both core stability and the necessary welfare bound. The essence of the lemma is rather simple; it says that if we take some project $k$ and add any arbitrary set of elements $T$ on top of $S_k$, then the total resulting welfare is no larger than the final payments to the agents in $T \cup S_k$. We first state the Doubling Lemmma here and then prove that using this lemma as a black-box, we can obtain both our welfare and stability result. The proof of the lemma is deferred to the Appendix.

\begin{lemma}
\label{lem_doublinglemma}
(Doubling Lemma) Consider any project $k$ and the set of elements assigned to $k$ in our solution ($S_k$). Let $T$ be some set of agents such that $T \cap S_k  = \emptyset$ and $|T| > |S_k|$. Then, the total payment to the agents in $T \cup S_k$ is at least $v_k(S_k \cup T)$, i.e.,
$$\sum_{i \in T \cup S_k} \bar{p}_i \geq v_k(S_k \cup T).$$
\end{lemma}

\subsection*{Proof of Core stability}
We need to show that our solution $(S, (\bar{p})_i)$ is core stable, i.e., for every project $k$ and set of agents $T$, $v_k(S_k \cup T) \leq \sum_{i \in S_k \cup T}\bar{p}_i$. Assume by contradiction that $\exists$ some project $k$ and some set $T$ that does not satisfy the inequality for stability. We claim that $|T| > |S_k|$.

\begin{lemma}
\label{lem_sizeoft}
If $v_k(S_k \cup T) > \sum_{i \in S_k \cup T}\bar{p}_i$, then there are strictly more agents in the set $T$ than in project $k$ under $S$, i.e., $|T| > |S_k|$.
\end{lemma}
\begin{subproof}
We know that
$$v_k(T \cup S_k) > \sum_{i \in T \cup S_k}\bar{p}_i \geq 2\sum_{i \in S_k}p_i = 2v_k(S_k).$$

Let $Q$ be some arbitrary set of size $\frac{|T \cup S_k|}{2}$. Applying Proposition~\ref{prop_anon_fundamental}, we get,

$$v_k(Q) \geq \frac{1}{2}v_k(T \cup S_k) > v_k(S_k).$$
By monotonicity, it must be that $|Q| > |S_k|$, and so $|T \cup S_k| > 2|S_k|$ giving us the desired lemma.  \end{subproof}

So, $S_k$ and $T$ satisfy the conditions required for the doubling lemma. Applying the lemma, we get $\sum_{i \in S_k \cup T}\bar{p}_i \geq v_k(S_k \cup T)$, which is a contradiction. Therefore, our solution is indeed core stable.

\subsection*{Welfare Bound}
Suppose that the optimum solution $O^* = (O^*_1, \ldots, O^*_M)$ has a social welfare of $v(O^*)$. We need to show that the social welfare of our solution $v(S)$ is at least half of $v(O^*)$. Recall that our social welfare is exactly equal to half the payments $\sum_i \bar{p}_i$. Therefore, it suffices if we prove that the welfare of the optimal solution $v(O^*)$ is not larger than the sum of the payments. Our approach is as follows: we will map every project $k$ to a proxy set $X_k \subseteq N$ so that $\sum_{i \in X_k}\bar{p}_i \geq v_k(O^*_k)$. If we ensure that the sets $(X_k)_{k \in \mathcal{P}}$ are mutually disjoint, we can sum these inequalities up to get our desired welfare bound.

We begin by dividing the projects into three categories based on the number of agents assigned to these projects in our solution $|S_k|$ and how this compares to $|O^*_k|$,
\begin{enumerate}
\item $(P_1)$ All projects $k$ satisfying, $|O^*_k| \geq |S_k| \geq \frac{1}{2} |O^*_k|$,
\item $(P_2)$ All projects $k$ satisfying $|S_k| > |O^*_k|$,
\item $(P_3)$ All projects $k$ satisfying $|O^*_k| > 2|S_k|$, i.e., in the optimum solution $k$ has more than double the number of agents assigned to $k$ in our solution.
\end{enumerate}

We define the sets $X_k$ as follows: for every project $k \in P_1$, $X_k = S_k$. For every project $k \in P_2$, $X_k$ is defined as the set of $|O^*_k|$ agents in $S_k$ with the highest payments as per our solution. Notice that for every $k \in P_2$, there are some `left over' agents who are not yet assigned to any $X_{k'}$. Let $Rem$ be the union of such leftover agents over all projects in $P_2$.

Finally, for every project $k \in P_3$, we define $X_k$ to be $S_k$ plus some arbitrarily chosen $|O^*_k| - |S_k|$ agents from the set $Rem$. It is not hard to see that we can choose $X_k$'s for the projects in $P_3$ in such a manner that these sets are all mutually disjoint. Indeed, this is true because
$$|Rem| + \sum_{k \in P_3} |S_k| \geq \sum_{k \in P_3}|O^*_k|.$$

The above inequality comes from the fact that $|O^*_k| - |X_k|$ summed over all $k \in P_1 \cup P_2$ is a non-negative number. Now all that remains is for us to show that $\sum_{i \in X_k}\bar{p}_i \geq v_k(O^*_k)$ for every $k$.

First, look at the projects in $P_1$. We can show that $\sum_{i \in X_k}\bar{p}_i = 2v_k(S_k) \geq 2.\frac{1}{2}v_k(O^*_k)$, where the last inequality comes from Proposition~\ref{prop_anon_fundamental} since $S_k$ has at least half as many agents as $O^*_k$. Now, for the projects in $P_2$, we can directly apply Lemma~\ref{lem_welfarepayupperb} to get $\sum_{i \in X_k} \bar{p}_i = 2\sum_{i \in X_k}p_i \geq v_k(O^*_k)$.

Finally, look at the projects in $P_3$. Fix some $k \in P_3$, and define $T = X_k \setminus S_k$. Since $|X_k| = |O^*_k|$, we immediately get $|T| > |S_k|$ from the definition of $P_3$. Therefore, we can apply the important Doubling Lemma and get the desired result. This completes the proof of our final welfare bound. \\

\end{mainproof}

\noindent\textbf{Envy-Free Payments} One interpretation for projects having anonymous valuations is that all the agents possess the same level of skill, and therefore, the value generated from a project depends only on the number of agents assigned to it. In such scenarios, it may be desirable that the payments given to the different agents are `fair' or envy-free, i.e., all agents assigned to a certain project must receive the same payment. The following theorem (which we formally prove in the Appendix) shows that Algorithm~\ref{alg_greedyanon} can be used to compute a $(2,2)$-approximate core that also satisfies this additional constraint of envy-freeness.

\begin{clm}
\label{clm_envyfreeanonymous}
For any instance where the projects have anonymous subadditive valuations, there exists a $(2,2)$-core stable solution such that the payments are envy-free, i.e., all the agents assigned to a single project receive the same payment.
\end{clm}

\subsection{Submodular and Fractionally Subadditive (XoS) Valuations}
Submodular and Fractionally Subadditive valuations are arguably the most popular classes of subadditive functions, and we show several interesting and improved results for these sub-classes. For instance, for XoS valuations, we can compute a $(1+\epsilon)$-core using Demand and XoS oracles (see~\cite{dobzinskiNS10} for a treatment of XoS oracles), whereas without these oracles, we can still compute a $(\frac{e}{e-1})$-core. For submodular valuations, we provide an algorithm to compute a $(1+\epsilon)$-core even without a Demand oracle. All of these solutions retain at least a fraction $(1-\frac{1}{e})$ of the optimum welfare, which matches the computational lower bound for both of these classes. We begin with a simple existence result for XoS valuations, that the optimum solution along with payments obtained using a XoS oracle form an exact core stable solution. All the results that we state in this Section, and Section~\ref{sec:auctions} are proved in the Appendix.

\begin{proposition}\label{prop:XoSexistence}
There exists a $(1,1)$-core stable solution for every instance where the projects have XoS valuations.
\end{proposition}

\noindent Since $Submodular \subset XoS$, this result extends to Submodular valuations as well. Unfortunately, it is known that the optimum solution cannot be computed efficiently for either of these classes unless P=NP~\cite{dobzinskiNS10}. However, we show that one can efficiently compute approximately optimal solutions that are almost-(core)-stable.

\begin{theorem}
\label{thm_mainxossubmod}
\begin{enumerate}
\item For any instance where the projects have XoS valuations, we can compute $(1+\epsilon,\frac{e}{e-1})$-core stable solution using Demand and XoS oracles, and a $(\frac{e}{e-1},\frac{e}{e-1})$-core stable solution without these oracles.

\item For submodular valuations, we can compute a $(1+\epsilon,\frac{e}{e-1})$-core stable solution using only a Value oracle.
\end{enumerate}
\end{theorem}
Note that for both the classes, a $(1+\epsilon)$-core can be computed in time polynomial in the input, and $\frac{1}{\epsilon}$. We conclude by pointing out that the results above are much better than what could have been obtained by plugging in $\alpha=\frac{e}{e-1}$ in Theorem~\ref{thm_mainsubadditive} for Submodular or XoS valuations.

\section{Relationship to Combinatorial Auctions}
\label{sec:auctions}
We now change gears and consider the seemingly unrelated problem of \emph{Item Bidding Auctions}, and establish a surprising equivalence between Core stable solutions and pure Nash equilibrium in Simultaneous Second Price Auctions. Following this, we adapt some of our results specifically for the auction setting and show how to efficiently compute Approximate Nash equilibrium when buyers have anonymous or submodular functions.

In recent years, the field of Auction Design has been marked by a paradigm shift towards `simple auctions'; one of the best examples of this is the growing popularity of Simultaneous Combinatorial Auctions~\cite{bhawalkarR11,christodoulouKS08,dobzinskiFK15}, where the buyers submit a single bid for each item. The auction mechanism is simple: every buyer submits one bid for each of the $N$ items, the auctioneer then proceeds to run $N$-parallel single-item auctions (usually first-price or second-price). In the case of Second Price Auctions, each item is awarded to the highest bidder (for that item) who is then charged the bid of the second highest bidder. Each buyer's utility is her valuation for the bundle she receives minus her total payment.

We begin by establishing that for every instance of our utility sharing problem, there is a corresponding combinatorial auction, and vice-versa. Formally, given an instance $(\mathcal{N}, \mathcal{P}, (v)_{k \in \mathcal{P}})$, we define the following `flipped auction': there is a set $\mathcal{N}$ of $N$ items, and a set $\mathcal{P}$ of $m$ buyers. Every buyer $k \in \mathcal{P}$ has a valuation function $v_k$ for the items. In the simultaneous auction, the strategy of every buyer is a bid vector $\vec{b_k}$; $b_k(i)$ denotes buyer $k$'s bid for item $i \in \mathcal{N}$. A profile of bid vectors $(\vec{b_1},\ldots,\vec{b_m})$ along with an allocation is said to be a pure Nash equilibrium  of the simultaneous auction if no buyer can unilaterally change her bids and improve her utility at the new allocation.

\subsubsection*{Over-Bidding} Nash equilibrium in Simultaneous Auctions is often accompanied by a rather strong \emph{no-overbidding} condition that a player's aggregate bid for every set $S$ of items is at most her valuation $v_k(S)$ for that set. In this paper, we also study the slightly less stringent \emph{weak no-overbidding} assumption considered in~\cite{fuKL12} and~\cite{feldmanFGL13} which states that `a player's total bid for her winning set is at most her valuation for that set'. The set of equilibrium with no-overbidding is strictly contained in the set of equilibrium with weak no-overbidding. Finally, to model buyers who overbid by small amounts, we focus on the following natural relaxation of no-overbidding known as $\gamma$-conservativeness that was defined by Bhawalkar and Roughgarden~\cite{bhawalkarR11}.

\begin{definition}{(Conservative Bids)~\cite{bhawalkarR11}}
For a given buyer $k \in \mathcal{P}$, a bid vector $\vec{b_k}$ is said to be $\gamma$-conservative if for all $T \subseteq \mathcal{N}$, we have $\sum_{i \in T}b_k(i) \leq \gamma\cdot v_k(T).$
\end{definition}

\noindent We now state our main equivalence result that is based on a simple black-box transformation to convert a Core stable solution $(S,\vec{\bar{p}})$ to a profile of bids $(\vec{b_k})_{k \in \mathcal{P}}$ that form a Nash Equilibrium: $b_k(i)=\bar{p}_i$ if $i \in S_k$, and $b_k(i)=0$ otherwise.

\begin{theorem}
\label{thm_equivalencecombauc}
Every Core stable solution for a given instance of our game can be transformed into a Pure Nash Equilibrium (with weak no-overbidding) of the corresponding `flipped' simultaneous second price auction, and vice-versa.
\end{theorem}

\noindent\textbf{Existence and Computation of Equilibrium} Although simultaneous auctions enjoy several desirable properties like good \emph{Price of Anarchy}~\cite{bhawalkarR11,christodoulouKS08}, their applicability is limited by both existential and computational barriers. Particularly, while a no over-bidding Nash equilibrium always exists for simple valuations like XoS, it may not be possible to actually compute one~\cite{dobzinskiFK15}. For more general subadditive (and even anonymous) valuations, Nash equilibria without over-bidding may not even exist~\cite{bhawalkarR11}, and whether or not they exist cannot be determined without exponential communication~\cite{dobzinskiFK15}. 

\noindent\textbf{A case for Approximate Equilibrium}
The exciting connection between Core stable solutions and Nash Equilibrium unfortunately extends to negative results as well. One can extend our lower bound examples (See Appendix) to show that even when all buyers have anonymous subadditive functions, there exist instances where every Nash equilibrium requires $O(\sqrt{N})$-conservative bids. The expectation that buyers will overbid by such a large amount appears to be unreasonable. In light of these impossibility results and the known barriers to actually compute a (no-overbidding) equilibrium~\cite{dobzinskiFK15}, we argue that in many auctions, it seems reasonable to consider $\alpha$-approximate Nash equilibrium that guarantee that buyers' utilities cannot improve by more than a factor $\alpha$ when they change their bids. In the following result, we adapt our previous algorithms to compute approximate equilibria with high social welfare for two useful settings. Moreover, these solutions require small over-bidding, and can be obtained via simple mechanisms, so it seems likely that they would actually arise in practice when pure equilibria either do not exist or require a large amount of overbidding. 

\begin{clm}
\label{clm_combauc_comp}
Given a Second Price Simultaneous Combinatorial Auction, we can compute in time polynomial in the input (and $\frac{1}{\epsilon}$ for a given $\epsilon > 0$)
\begin{enumerate}
\item A $2$-approximate Nash equilibrium that extracts half the optimal social welfare as long as the buyers have anonymous subadditive valuations.
\item A $(1+\epsilon)$-approximate Nash equilibrium that is a $\frac{e}{e-1}$-approximation to the optimum welfare when the buyers have submodular valuations.

\end{enumerate}

The first solution involves $4$-conservative bids, and the second solution involves $(1+\epsilon)$-conservative bids.
\end{clm}

Given a submodular valuation $v_k$, define $v_{max} = \max_{i, S} v_k(i | S)$. Also, define $\Delta = \min_{i,S}v_k(i|S)$ such that $v_k(i|S) > 0$. That is $\Delta$ is the smallest non-zero increment in utility. Then, the algorithm for Submodular Functions converges in Poly($N,m,\frac{1}{\epsilon}, \log(\frac{v_{max}}{\Delta}))$ time. One can contrast this result to an algorithm by~\cite{dobzinskiFK15} that computes an exact Nash equilibrium in pseudo-polynomial time, i.e., $O(\frac{v_{max}}{\Delta})$. On the contrary, we show that we can compute an approximate Nash equilibrium in poly-time (using a PTAS).

\subsection*{Conclusion} We conclude by remarking that despite the large body of work in Simultaneous Auctions, our main results do not follow from any known results in that area, and we hope that our techniques lead to new insights for computing auction equilibria. 



\bibliography{bibliography}

\begin{thebibliography}{10}

\bibitem{anshelevichS14}
Elliot Anshelevich and Shreyas Sekar.
\newblock Approximate equilibrium and incentivizing social coordination.
\newblock In {\em Proceedings of the Twenty-Eighth {AAAI} Conference on
  Artificial Intelligence, July 27 -31, 2014, Qu{\'{e}}bec City, Qu{\'{e}}bec,
  Canada.}, pages 508--514, 2014.

\bibitem{augustineCEFGS15}
John Augustine, Ning Chen, Edith Elkind, Angelo Fanelli, Nick Gravin, and
  Dmitry Shiryaev.
\newblock Dynamics of profit-sharing games.
\newblock {\em Internet Mathematics}, 11(1):1--22, 2015.

\bibitem{aumann1964bargaining}
Robert~J Aumann and Michael Maschler.
\newblock The bargaining set for cooperative games.
\newblock {\em Advances in game theory}, 52:443--476, 1964.

\bibitem{bachrach09cost}
Yoram Bachrach, Edith Elkind, Reshef Meir, Dmitrii~V. Pasechnik, Michael
  Zuckerman, J{\"{o}}rg Rothe, and Jeffrey~S. Rosenschein.
\newblock The cost of stability in coalitional games.
\newblock In {\em Algorithmic Game Theory, Second International Symposium,
  {SAGT} 2009, Paphos, Cyprus, October 18-20, 2009. Proceedings}, pages
  122--134, 2009.

\bibitem{bachrachPR13}
Yoram Bachrach, David~C. Parkes, and Jeffrey~S. Rosenschein.
\newblock Computing cooperative solution concepts in coalitional skill games.
\newblock {\em Artif. Intell.}, 204:1--21, 2013.

\bibitem{balcanBM09}
Maria{-}Florina Balcan, Avrim Blum, and Yishay Mansour.
\newblock Improved equilibria via public service advertising.
\newblock In {\em Proceedings of the Twentieth Annual {ACM-SIAM} Symposium on
  Discrete Algorithms, {SODA} 2009, New York, NY, USA, January 4-6, 2009},
  pages 728--737, 2009.

\bibitem{bejanG09}
Camelia Bejan and Juan~Camilo G{\'{o}}mez.
\newblock Core extensions for non-balanced tu-games.
\newblock {\em Int. J. Game Theory}, 38(1):3--16, 2009.

\bibitem{bhalgatCK10}
Anand Bhalgat, Tanmoy Chakraborty, and Sanjeev Khanna.
\newblock Approximating pure nash equilibrium in cut, party affiliation, and
  satisfiability games.
\newblock In {\em Proceedings 11th {ACM} Conference on Electronic Commerce
  (EC-2010), Cambridge, Massachusetts, USA, June 7-11, 2010}, pages 73--82,
  2010.

\bibitem{bhawalkarR11}
Kshipra Bhawalkar and Tim Roughgarden.
\newblock Welfare guarantees for combinatorial auctions with item bidding.
\newblock In {\em Proceedings of the Twenty-Second Annual {ACM-SIAM} Symposium
  on Discrete Algorithms, {SODA} 2011, San Francisco, California, USA, January
  23-25, 2011}, pages 700--709, 2011.

\bibitem{vetta2015coalition}
Nicolas Bousquet, Zhentao Li, and Adrian Vetta.
\newblock Coalition games on interaction graphs: A horticultural perspective.
\newblock In {\em Proceedings of the Sixteenth ACM Conference on Economics and
  Computation}, EC '15, pages 95--112, New York, NY, USA, 2015. ACM.

\bibitem{branzeiL09}
Simina Br{\^{a}}nzei and Kate Larson.
\newblock Coalitional affinity games and the stability gap.
\newblock In {\em {IJCAI} 2009, Proceedings of the 21st International Joint
  Conference on Artificial Intelligence, Pasadena, California, USA, July 11-17,
  2009}, pages 79--84, 2009.

\bibitem{chalkiadakisEMPJ10}
Georgios Chalkiadakis, Edith Elkind, Evangelos Markakis, Maria Polukarov, and
  Nick~R. Jennings.
\newblock Cooperative games with overlapping coalitions.
\newblock {\em J. Artif. Intell. Res. {(JAIR)}}, 39:179--216, 2010.

\bibitem{chierichettiKO13}
Flavio Chierichetti, Jon~M. Kleinberg, and Sigal Oren.
\newblock On discrete preferences and coordination.
\newblock In {\em {ACM} Conference on Electronic Commerce, {EC} '13,
  Philadelphia, PA, USA, June 16-20, 2013}, pages 233--250, 2013.

\bibitem{christodoulouKS08}
George Christodoulou, Annam{\'{a}}ria Kov{\'{a}}cs, and Michael Schapira.
\newblock Bayesian combinatorial auctions.
\newblock In {\em Automata, Languages and Programming, 35th International
  Colloquium, {ICALP} 2008, Reykjavik, Iceland, July 7-11, 2008, Proceedings,
  Part {I:} Tack {A:} Algorithms, Automata, Complexity, and Games}, pages
  820--832, 2008.

\bibitem{deng1999algorithmic}
Xiaotie Deng, Toshihide Ibaraki, and Hiroshi Nagamochi.
\newblock Algorithmic aspects of the core of combinatorial optimization games.
\newblock {\em Mathematics of Operations Research}, 24(3):751--766, 1999.

\bibitem{devanurMV05}
Nikhil~R. Devanur, Milena Mihail, and Vijay~V. Vazirani.
\newblock Strategyproof cost-sharing mechanisms for set cover and facility
  location games.
\newblock {\em Decision Support Systems}, 39(1):11--22, 2005.

\bibitem{dobzinskiFK15}
Shahar Dobzinski, Hu~Fu, and Robert~D. Kleinberg.
\newblock On the complexity of computing an equilibrium in combinatorial
  auctions.
\newblock In {\em Proceedings of the Twenty-Sixth Annual {ACM-SIAM} Symposium
  on Discrete Algorithms, {SODA} 2015, San Diego, CA, USA, January 4-6, 2015},
  pages 110--122, 2015.

\bibitem{dobzinskiNS10}
Shahar Dobzinski, Noam Nisan, and Michael Schapira.
\newblock Approximation algorithms for combinatorial auctions with
  complement-free bidders.
\newblock {\em Math. Oper. Res.}, 35(1):1--13, 2010.

\bibitem{fangKZ08}
Qizhi Fang, Liang Kong, and Jia Zhao.
\newblock Core stability of vertex cover games.
\newblock {\em Internet Mathematics}, 5(4):383--394, 2008.

\bibitem{feige09subadd}
Uriel Feige.
\newblock On maximizing welfare when utility functions are subadditive.
\newblock {\em {SIAM} J. Comput.}, 39(1):122--142, 2009.

\bibitem{feldmanF15}
Michal Feldman and Ophir Friedler.
\newblock A unified framework for strong price of anarchy in clustering games.
\newblock In {\em Automata, Languages, and Programming - 42nd International
  Colloquium, {ICALP} 2015, Kyoto, Japan, July 6-10, 2015, Proceedings, Part
  {II}}, pages 601--613, 2015.

\bibitem{feldmanFGL13}
Michal Feldman, Hu~Fu, Nick Gravin, and Brendan Lucier.
\newblock Simultaneous auctions are (almost) efficient.
\newblock In {\em Symposium on Theory of Computing Conference, STOC'13, Palo
  Alto, CA, USA, June 1-4, 2013}, pages 201--210, 2013.

\bibitem{feldmanLN12}
Moran Feldman, Liane Lewin{-}Eytan, and Joseph Naor.
\newblock Hedonic clustering games.
\newblock In {\em 24th {ACM} Symposium on Parallelism in Algorithms and
  Architectures, {SPAA} '12, Pittsburgh, PA, USA, June 25-27, 2012}, pages
  267--276, 2012.

\bibitem{fuKL12}
Hu~Fu, Robert Kleinberg, and Ron Lavi.
\newblock Conditional equilibrium outcomes via ascending price processes with
  applications to combinatorial auctions with item bidding.
\newblock In {\em {ACM} Conference on Electronic Commerce, {EC} '12, Valencia,
  Spain, June 4-8, 2012}, page 586, 2012.

\bibitem{georgiou2013black}
Konstantinos Georgiou and Chaitanya Swamy.
\newblock Black-box reductions for cost-sharing mechanism design.
\newblock {\em Games and Economic Behavior}, 2013.

\bibitem{goemans2004cooperative}
Michel~X Goemans and Martin Skutella.
\newblock Cooperative facility location games.
\newblock {\em Journal of Algorithms}, 50(2):194--214, 2004.

\bibitem{grecoMPS11}
Gianluigi Greco, Enrico Malizia, Luigi Palopoli, and Francesco Scarcello.
\newblock On the complexity of the core over coalition structures.
\newblock In {\em {IJCAI} 2011, Proceedings of the 22nd International Joint
  Conference on Artificial Intelligence, Barcelona, Catalonia, Spain, July
  16-22, 2011}, pages 216--221, 2011.

\bibitem{hoefer13strategic}
Martin Hoefer.
\newblock Strategic cooperation in cost sharing games.
\newblock {\em Int. J. Game Theory}, 42(1):29--53, 2013.

\bibitem{immorlicaMM08}
Nicole Immorlica, Mohammad Mahdian, and Vahab~S. Mirrokni.
\newblock Limitations of cross-monotonic cost-sharing schemes.
\newblock {\em {ACM} Transactions on Algorithms}, 4(2), 2008.

\bibitem{kleinbergO11}
Jon~M. Kleinberg and Sigal Oren.
\newblock Mechanisms for (mis)allocating scientific credit.
\newblock In {\em Proceedings of the 43rd {ACM} Symposium on Theory of
  Computing, {STOC} 2011, San Jose, CA, USA, 6-8 June 2011}, pages 529--538,
  2011.

\bibitem{lehmannLN06}
Benny Lehmann, Daniel~J. Lehmann, and Noam Nisan.
\newblock Combinatorial auctions with decreasing marginal utilities.
\newblock {\em Games and Economic Behavior}, 55(2):270--296, 2006.

\bibitem{lewenbergBSZR15}
Yoad Lewenberg, Yoram Bachrach, Yonatan Sompolinsky, Aviv Zohar, and Jeffrey~S.
  Rosenschein.
\newblock Bitcoin mining pools: {A} cooperative game theoretic analysis.
\newblock In {\em Proceedings of the 2015 International Conference on
  Autonomous Agents and Multiagent Systems, {AAMAS} 2015, Istanbul, Turkey, May
  4-8, 2015}, pages 919--927, 2015.

\bibitem{lucierB10}
Brendan Lucier and Allan Borodin.
\newblock Price of anarchy for greedy auctions.
\newblock In {\em Proceedings of the Twenty-First Annual {ACM-SIAM} Symposium
  on Discrete Algorithms, {SODA} 2010, Austin, Texas, USA, January 17-19,
  2010}, pages 537--553, 2010.

\bibitem{markakis2005core}
Evangelos Markakis and Amin Saberi.
\newblock On the core of the multicommodity flow game.
\newblock {\em Decision support systems}, 39(1):3--10, 2005.

\bibitem{meirBR10}
Reshef Meir, Yoram Bachrach, and Jeffrey~S. Rosenschein.
\newblock Minimal subsidies in expense sharing games.
\newblock In {\em Algorithmic Game Theory - Third International Symposium,
  {SAGT} 2010, Athens, Greece, October 18-20, 2010. Proceedings}, pages
  347--358, 2010.

\bibitem{moulin1999incremental}
Herv{\'e} Moulin.
\newblock Incremental cost sharing: Characterization by coalition
  strategy-proofness.
\newblock {\em Social Choice and Welfare}, 16(2):279--320, 1999.

\bibitem{moulin2001strategyproof}
Herv{\'e} Moulin and Scott Shenker.
\newblock Strategyproof sharing of submodular costs: budget balance versus
  efficiency.
\newblock {\em Economic Theory}, 18(3):511--533, 2001.

\bibitem{myerson1977graphs}
Roger~B Myerson.
\newblock Graphs and cooperation in games.
\newblock {\em Mathematics of operations research}, 2(3):225--229, 1977.

\bibitem{roughgardenS09}
Tim Roughgarden and Mukund Sundararajan.
\newblock Quantifying inefficiency in cost-sharing mechanisms.
\newblock {\em J. {ACM}}, 56(4), 2009.

\bibitem{saad2009coalitional}
Walid Saad, Zhu Han, M{\'e}rouane Debbah, Are Hj{\o}rungnes, and Tamer
  Ba{\c{s}}ar.
\newblock Coalitional game theory for communication networks.
\newblock {\em Signal Processing Magazine, IEEE}, 26(5):77--97, 2009.

\bibitem{schmeidler1969nucleolus}
David Schmeidler.
\newblock The nucleolus of a characteristic function game.
\newblock {\em SIAM Journal on applied mathematics}, 17(6):1163--1170, 1969.

\bibitem{schulzU13}
Andreas~S. Schulz and Nelson~A. Uhan.
\newblock Approximating the least core value and least core of cooperative
  games with supermodular costs.
\newblock {\em Discrete Optimization}, 10(2):163--180, 2013.

\bibitem{shapley1966quasi}
Lloyd~S Shapley and Martin Shubik.
\newblock Quasi-cores in a monetary economy with nonconvex preferences.
\newblock {\em Econometrica: Journal of the Econometric Society}, pages
  805--827, 1966.

\bibitem{vondrak2007submodularity}
Jan Vondr{\'a}k.
\newblock {\em Submodularity in combinatorial optimization}.
\newblock PhD thesis, Citeseer, 2007.

\bibitem{vondrak08}
Jan Vondr{\'{a}}k.
\newblock Optimal approximation for the submodular welfare problem in the value
  oracle model.
\newblock In {\em Proceedings of the 40th Annual {ACM} Symposium on Theory of
  Computing, Victoria, British Columbia, Canada, May 17-20, 2008}, pages
  67--74, 2008.

\end{thebibliography}
\bibliographystyle{plain}

\appendix

\newpage
\section{Appendix: Proofs for Subadditive Valuations}
\label{app:mainsub}
We begin by formally defining the Greedy Matching Procedure procedure that is the building block of our main algorithm.

\begin{algorithm}
    \SetKwInOut{Input}{Input}
    \SetKwInOut{Output}{Output}

    \Input{Allocation $I=(I_1, \ldots, I_M)$, Payments $\vec{p^I}=(p^I_1, \ldots, p^I_N)$.}
    \Output{Allocation $O=(O_1, \ldots, O_M)$, Payments $\vec{p^O} = (p^O_1, \ldots, p^O_N)$.}
    Initialize the current allocation $S=I$, and current payments $\vec{p}=\vec{p^I}$. \\
    \eIf{$\exists$ empty project $k$ with $S_k = \emptyset$ and agent $i$ such that $v_k(i) > p_i$}
      {
        Remove agent $i$ from her current project and assign her to the empty project $l$ with the maximum value of $v_l(i)$.
        Update agent $i$'s payment to $p_i = v_l(i)$.
      }
      {
        return the current allocation $S$ and payments $\vec{p}$.
      }
    \caption{Greedy Matching with Reserve Prices}
    \label{alg_greedy}
\end{algorithm}

\begin{lem_app}{lem_subadd_halfagentsgone}
For every bad project $k$, $|A^+_k| > |A^-_k|$, i.e., more than half the agents in $A_k$ still remain in project $k$.

\end{lem_app}
\begin{subproof}

We prove this by contradiction. Suppose that for some such $k$, $|A^-_k| \geq |A^+_k|$. Recall that the marginal contributions given to agents in $A^-_k$ is
$$\sum_{i \in A^-_k}p^*_i + \frac{|A^-_k|}{|A_k|}z^*_k \geq \sum_{i \in A^-_k}p^*_i + \frac{z^*_k}{2} \quad \quad (\text{since $A_k = A^+_k \cup A^-_k$}).$$
Moreover, the payments are non-decreasing under Algorithm~\ref{alg_greedy}, and agent $i \in A^-_k$ is transferred from project $k$ to some project $l$ only if $v_l(i) > p^*_i + \frac{z^*_k}{|A_k|}$. Therefore, we have,
$$ \sum_{l \in P_k}v_l(B_l)  > \sum_{i \in A^-_k}(p^*_i + \frac{z^*_k}{|A_k|}) \geq \sum_{i \in A^-_k}p^*_i + \frac{z^*_k}{2}.$$
Now, $\vec{p^*}$ and $\vec{z^*}$ are feasible solutions to the dual LP; this means that $\sum_{i \in A^-_k}p^*_i + z^*_k \geq v_k(A^-_k)$. So, we have,
\begin{align*}
v_k(A^+_k) + \sum_{l \in P_k}v_l(B_l) & \geq v_k(A^+_k) + \sum_{i \in A^-_k}p^*_i + \frac{z^*_k}{2}\\
& > v_k(A^+_k) + \frac{1}{2}v_k(A^-_k)\\
& \geq \frac{1}{2} (v_k(A^+_k) + v_k(A^-_k))\\
& \geq \frac{1}{2} v_k(A_k).
\end{align*}
The last inequality follows from subadditivity and the fact that $A_k = A^+_k \cup A^-_k$. However, this contradicts the definition of a \emph{bad} project. So, we must have that $|A^-_k| < |A^+_k|.$
\end{subproof}


\subsection*{Main Algorithm - Phase I}
\begin{enumerate}
\item Run Algorithm~\ref{alg_greedy} on the allocation $A$ with payments $\vec{p^0}$. Let the output be $B, \vec{p^B}$, and define $A^+_k$, $A^-_k$, and $P_k$ as mentioned above.

\item (Good Projects) For every \emph{good} project $k \in \mathcal{P} \setminus \zeta(A)$
\begin{enumerate}
\item $\forall i \in A^+_k$, assign them to project $k$, set $p'_i = p^*_i + \frac{z^*_k}{|A^+_k|}$
\item $\forall i \in A^-_k$, assign them to their project $l\in P_k$, set $p'_i = p^*_i + z^*_l$
\item Denote the resulting sets by $S'_k$ and $S'_l$
\end{enumerate}

\item (Bad Projects) For every \emph{bad} project $k \in \mathcal{P} \setminus \zeta(A)$
\begin{enumerate}
\item Arbitrarily choose dummy agents $D_k \subset A^+_k$ such that $|D_k| = |A^+_k| - |A^-_k|$ (this is possible due to Lemma~\ref{lem_subadd_halfagentsgone}).

\item Set $S'_k = A^-_k \cup D_k$.

\item For $i \in A^-_k$, set $p'_i = p^B_i + \frac{z'_k}{|A^-_k|}$ where $z'_k$ is the leftover slack defined as $z^*_k - \sum_{i \in A^-_k}(p^B_i - p^*_i)$. \\
For $i \in D_k$, set $p'_i = p^*_i$.

\item Assign the non-dummy agents in $A^+_k$ arbitrarily to the projects in $P_k$ so that each project in $P_k$ gets exactly one agent from $(A^+_k \setminus D_k)$. ($S'_l$ is defined accordingly for $l \in P_k$).

\item For every $l \in P_k$: for $i \in S'_l$, set $p'_i = p^*_i + z^*_l$.
\end{enumerate}
\end{enumerate}

%

\begin{lem_app}{lem_subadd_nondummy}
For every agent $i$ that does not belong to the set of dummy agents, her payment at the end of the first phase ($p'_i$) is at least her payment returned by the call to the Greedy Matching Procedure $p^B_i$.
\end{lem_app}
\begin{subproof}
For each agent assigned to a \emph{good} project $k$ and the associated set of agents in $P_k$, the claim follows almost trivially from Lemma~\ref{lem_match_upperbound} and the definition of the final payment $p'_i$. Moreover, suppose that $k$ is a \emph{bad} project, then we claim that the leftover slack $z'_k$ is non-negative. This is true because
$$\sum_{i \in A^-_k}p^B_i = \sum_{l \in P_k}v_l(B_l) < v_k(A_k)/2 < v_k(A^-_k) \leq \sum_{i \in A^-_k}p^*_i + z^*_k.$$

(The first inequality above is since $k$ is a bad project; the second is due to the fact that $v_k$ is subadditive and $v_k(A_k^+)<v_k(A_k)/2$ since $k$ is a bad project. The last is due to dual feasibility.)

Therefore, by definition, the payment to a non-dummy agent $i$ assigned to a bad project $p'_i = p^B_i + \frac{z'_l}{|A^-_k|} \geq p^B_i$. Finally, fix a \emph{bad} project $k$ and consider the agents assigned to the projects in $P_k$. We claim that for every $l \in P_k$, $z^*_l$ is at least as large as $\frac{z^*_k}{|A_k|}$. We first show how this claim leads to the desired lemma and then conclude by proving the claim. For any $l \in P_k$, let $S'_l$ be some agent $i$. Since agent $i\in A_k^+$, it remained on the same project $k$ in solution $B$. This means that her payment as output by the matching procedure is exactly the same as the input payment $p^*_i + \frac{z^*_k}{|A_k|}$, which by the above claim is not larger than $p^*_i + z^*_l$ giving us the desired lemma.

(Proof of Claim that $z^*_l \geq \frac{z^*_k}{|A_k|}$) Suppose that some agent $j \in A_k$ belonged to project $l$ in the solution returned by the matching $B$. Then, by the monotonicity of payments in Algorithm~\ref{alg_greedy}, it must hold that $j$'s payment at the termination of Algorithm~\ref{alg_greedy}, $p^B_j$, is at least her initial payment $p^*_j + \frac{z^*_k}{|A_k|}$. Applying dual feasibility, we get the desired claim

$$ p^*_j + z^*_l \geq v_l(j) = p^B_j \geq p^*_j + \frac{z^*_k}{|A_k|}.$$ \end{subproof}

\begin{lem_app}{lem_paymentfirstphase}
For every non-empty project $k \notin \zeta(S')$, the total payment to agents in $k$ at the end of Phase I is exactly $\sum_{i \in S'_k}p^*_i + z^*_k$.
\end{lem_app}
\begin{subproof}
First consider any \emph{good} project $k$ and the associated set of projects in $P_k$. By definition, for every $l \in k \cup P_k$, the total payments to the agents in $S'_l$ is exactly as given by the lemma. So the proof for \emph{good} projects and associated projects is trivial. Suppose that $k$ is a \emph{bad} project, then every $l \in P_k$ contains exactly one agent $i$ whose payment is exactly $p^*_i + z^*_l$. Finally, look at project $k$ and recall that $D_k$ is the set of dummy agents in $k$. Then, by definition of the payments to the agents in $S'_k$, we have
$$\sum_{i \in S'_k}p'_i = \sum_{i \in S'_k \setminus D_k}p'_i + \sum_{i \in D_k}p'_i = \sum_{i \in S'_k \setminus D_k}p^*_i + z^*_k + \sum_{i \in D_k}p^*_i.$$
\end{subproof}

%
%

\subsection*{Main Algorithm - Phase II}

\begin{enumerate}
\item Input to this Phase is the output of Phase I: $(S', \vec{p'})$
\item Run Algorithm~\ref{alg_greedy} with the input $(S', \vec{p'})$
\item Let the output of Algorithm~\ref{alg_greedy} be $(S,\vec{p}^{temp})$.
\item Construct the final payment vector $(\bar{p})_i$ as follows: if agent $i$ belongs to strategy $k$ in both $S$ and $S'$, then set $\bar{p}_i = p^{temp}_i$. Otherwise if agent $i$'s strategy in $S$ is $k$ but her strategy is $S'$ is not $k$, then set $\bar{p}_i = p^*_i + z^*_k$.
\item Output the final solution: $S$, $(\bar{p})_i$.
\end{enumerate}

We now prove some simple properties that compare the output of Phase II with its input.
\begin{clm_app}{clm_phase2invariants}
The following properties are true:
\begin{enumerate}
\item The set of empty projects in $S$ is a subset of the set of empty projects in $S'$, i.e., $\zeta(S) \subseteq \zeta(S')$.
\item For all non-dummy agents, their strategies in $S'$ and $S$ coincide.
\item For every project $k$ that was empty in $S'$ but not in $S$, $S_k$ consists of a single dummy agent $i$.
\item For every agent $i \in N$, her payment at the end of Phase II ($\bar{p}_i$) is at least her payment at the end of Phase I.
\end{enumerate}
\end{clm_app}
\begin{subproof}
Recall that in every stage of our greedy matching procedure, an agent $i$ is transferred to an empty group (say $k$) having the largest value of $v_k(i)$ as long as it is greater than the agent's current payment. Consider the execution of the greedy matching procedure in our algorithm's second phase. We prove by induction that at every stage of this procedure, (i) the set of empty projects is a subset of $\zeta(S)$, ii) for every non-dummy agent, her strategy remains the same as in $S'$. Clearly, this is true at the beginning. Moreover, note that every project $k \notin \zeta(S')$ contains at least one non-dummy agent under $S$'.

Suppose that the two induction hypotheses hold up to some iteration $t$ of the algorithm. Let $P_e$ be the set of empty projects at the end of this iteration, and let $\vec{p}$ be the payment vector after iteration $t$. In iteration $t+1$, the agent $i$ assigned to an empty project $k$ satisfies $p_i < v_k(i)$. This implies that $i$ cannot be a non-dummy agent because due to Corollary~\ref{corr_subadd_emptyproj} and the monotonicity of payments, we know that for every non-dummy agent $j$, $p_j \geq p'_j \geq v_k(i)$ since $k \in \zeta(S')$. Therefore, project $k$ gains a dummy agent in iteration $t+1$ and the positions of non-dummy agents remain the same as they were in the previous iteration. Moreover,  every non-empty project in $S'$ still has one non-dummy agent and therefore, remains non-empty. We conclude the proof of Properties $(1)$ and $(2)$. Property $(3)$ is simply a corollary of Property $(2)$.

Finally, we know by the monotonicity of payments that $\vec{p}^{temp} \geq \vec{p'}$. Moreover, for every agent $i$ whose strategy did not change from $S'$, her payment remains the same as in $\vec{p}^{temp}$ and therefore, the input payments. For any  dummy agent (say $i$) who transferred to an empty project (say $k$), her final payment is $p^*_i + z^*_k$ which by dual feasibility is not smaller than $p^{temp}_i = v_k(i)$ which in turn is larger than $p'_i$. Therefore, Property $(4)$ also holds.
\end{subproof}

\begin{lem_app}{lem_paymentbounds}
For every non-empty project $k \notin \zeta(S)$, the total payments made to agents in $k$ is exactly $\sum_{i \in S_k}p^*_i + z^*_k$. Moreover, the payment made to any agent $i$ is at least her dual price $p^*_i$.
\end{lem_app}
\begin{subproof}
The proof is rather straightforward and is analogous to Lemma~\ref{lem_paymentfirstphase}. The only change caused by Phase II is in the strategies of dummy agents, and because dummy agents only belong to $\emph{bad}$ projects, it suffices to show this lemma for $\emph{bad}$ projects and the projects that newly gained a dummy agent in Phase II, i.e., $k \in \zeta(S') \setminus \zeta(S)$. For the latter case, the lemma is true trivially due to the definition of Phase II. Now, consider some \emph{bad} project $k$, and let $X$ be the set of dummy agents belonging to $S_k$ (ones that did not deviate in the greedy matching procedure). Moreover, for every agent in $S_k$, her payment is the same as her input payment because these agents did not deviate during the course of the matching procedure. Therefore, the final payments to the agents can be aggregated similar to the method in Lemma~\ref{lem_paymentfirstphase},

$$\sum_{i \in S_k}\bar{p}_i = \sum_{i \in S_k \setminus X}\bar{p}_i + \sum_{i \in X}\bar{p}_i = \sum_{i \in S_k}p^*_i + z^*_k + \sum_{i \in X}p^*_i.$$

\end{subproof}

\section{Appendix: Proofs for Anonymous Subadditive Valuations}
\label{app:anon}
We begin with a basic property of anonymous subadditive functions that we require in all of the proofs. In the rest of this paper, the notation $v(T | S)$ refers to $v(S \cup T) - v(S)$.

\begin{proposition}
\label{prop_anon_fundamental}
Let $S \subseteq N$ be some set of agents, and suppose $T \subseteq N$ such that $|T| \geq \frac{|S|}{2}$. Then for an anonymous subadditive function $v$, we have $v(T) \geq \frac{v(S)}{2}$.
\end{proposition}
The proof follows from the fact that $v(T) + v(S \setminus T) \geq v(S)$, and $|T| \geq |S|$.

\begin{clm_app}{clm_badexamplesubadd}
\textbf{(Lower Bounds)} There exists instances having only two projects with anonymous subadditive functions such that
\begin{enumerate}

\item For any  $\epsilon > 0$, no $(2-\epsilon, c)$-core stable solution exists for any value $c$.

\item For any  $\epsilon > 0$, no $(\alpha, 2-\epsilon)$-core stable solution exists for any constant $\alpha$.
\end{enumerate}
\end{clm_app}

\begin{subproof}
%

(Part 2) Consider an instance with $N$ agents and $2$ projects. We choose a large enough $N$ so that the following condition is satisfied $\frac{N}{N/2 + \sqrt{N}} > 2-\epsilon$. Now the project valuations are defined as follows:
\begin{align*}
v_1(S) = & \frac{N}{2} \quad \text{for }S \neq \mathcal{N} & v_2(S) =  & \sqrt{N} \quad \forall S \subseteq \mathcal{N} \\
v_1(\mathcal{N}) = & N, & &
\end{align*}
The social welfare is maximized when all agents are allocated to project $1$. It is also not hard to see (due to our choice of $N$) that no other solution has a social welfare that is at most a factor $2-\epsilon$ away from OPT. Now, using the same reasoning as we did for the previous claim, we can conclude that in order to stabilize the optimum solution, every agent needs a payment of at least $\sqrt{N}$, and therefore, the total payments have to be at least a factor $\sqrt{N}$ larger than the optimum welfare.
\end{subproof}

\begin{lem_app}{lem_ano_monoton}
For every $i,j$ with $i < j$, the marginal of the agents in $A_i$ is not smaller than the marginal of the agents in $A_j$, i.e., $p^{(1)} \geq p^{(2)} \geq \ldots \geq p^{(r)}$.
\end{lem_app}
\begin{subproof}
We argue that it is sufficient if we prove the lemma only for the case where the agents in $A_i$ and $A_j$ are assigned to the same project in $S$. To see why, suppose that the lemma holds for every such pair, now look at some $A_i$, $A_j$ with $i < j$ such that the agents in $A_i$ belong to project $k_i$ and $A_j$ belong to project $k_j$ with $k_i \neq k_j$. Next, define $i < l \leq j$ to be the smallest index such that the agents in $A_l$ are assigned to project $k_j$, the same as the agents in $A_j$. Note that by definition of $l$, $S^{(i)}_{k_j} = S^{(l)}_{k_j}$: the set of agents in project $k_j$ are the same before iterations $i$, and $l$. This implies that the \emph{assignment of the agents in $A_l$ to project $k_j$} was a possible step for the greedy algorithm for iteration $i$. However, since the greedy algorithm actually chose $A_i \to k_i$, this means that

$$p^{(i)} = \frac{v_{k_i}(A_i | S^{(i)}_{k_i})}{|A_i|} \geq \frac{v_{k_j}(A_l | S^{(l)}_{k_j})}{|A_l|} = p^{(l)}.$$

But we know that $p^{(l)} \geq p^{(j)}$ since the agents in $A_l$ and $A_j$ are assigned to the same project. Therefore, it suffices to prove the lemma for the case where the two sets of agents are assigned to the same project, i.e., we need to prove that if the agents in $A_i$ and $A_j$ are assigned to the same project with $i < j$, then $p^{(i)} \geq p^{(j)}$.

Suppose that the agents in $A_i$ and $A_j$ are both assigned to project $k$ and assume by contradiction that $p^{(j)} > p^{(i)}$. Without loss of generality, we can assume that no agents are assigned to project $k$ in between iterations $i$ and $j$. Since $p^{(i)} < p^{(j)}$, we have
$$\frac{v_k(A_i | S^{(i)}_k)}{|A_i|} < \frac{v_k(A_j | S^{(i)}_k \cup A_i)}{|A_j|}.$$

Consider the set $A_i \cup A_j$. We have that
$$\frac{v_k(A_i \cup A_j | S^{(i)}_k )}{|A_i| + |A_j|} =  \frac{v_k(A_j | A_i \cup S^{(i)}_k) + v_k(A_i | S^{(i)}_k)}{|A_j| + |A_i|} > \frac{v_k(A_i | S^{(i)}_k)}{|A_i|}. $$

The last inequality follows from the basic algebraic property that $\frac{a+c}{b+d} > \min(\frac{a}{b}, \frac{c}{d})$ as long as $\frac{a}{b} \neq \frac{c}{d}.$ In other words, the average welfare due to adding the agents in $A_i \cup A_j$ to project $k$ during the $i^{th}$ iteration is strictly larger than the average welfare due to adding the agents in $A_i$ to project $k$ in the same iteration. However, this means that in iteration $i$, the algorithm would have chosen the agents in $A_i \cup A_j$ and assigned them to to project $k$ instead of the agents in $A_i$, which is a contradiction. This completes the proof.
\end{subproof}

\begin{lem_app}{lem_welfarepayupperb}
For every project $k$, and any given positive integer $t \leq |S_k|$, the total marginal contribution of the $t$ highest paid agents in $S_k$ is at least the value derived due to any set of $t$ agents, i.e., if $T$ denotes the set of $t$-highest paid agents in $S_k$, then

$$\sum_{i \in T}p_i \geq v_k(T).$$
\end{lem_app}
\begin{subproof}
Define $l$ to be the smallest index such that after the $l^{th}$ iteration, the project $k$ contains at least $t$ agents, i.e., $|S^{(l)}_k| < t$ and $|S^{(l+1)}_k| \geq t$. Then, from the monotonicity of the marginals (Lemma~\ref{lem_ano_monoton}) it is not hard to see that the set $T$ of the $t$ highest paid agents must contain $S^{(l)}_k$ and some arbitrary $t - |S^{(l)}_k|$ agents from the set of agents added in $l^{th}$ iteration. Consider the average welfare increase obtained by adding the agents in $T \setminus S^{(l)}_k$ to project $k$ during iteration $l$; this cannot be larger than $p^{(l)}$, the actual average welfare obtained by our greedy algorithm in that iteration, i.e.,

$$p^{(l)} \geq \frac{v_k(T) - v_k(S_k^{(l)})}{|T| - |S_k^{(l)}|}.$$

Therefore, the total marginal contributions of the $t$ top agents in $S_k$ can be bounded as follows,
$$\sum_{i \in T}p_i = v_k(S^{(l))}_k) + (|T| - |S_k^{(l)}|) p^{(l)} \geq v_k(T).$$

The first part of the above inequality comes from the fact that the total payment to the agents in $S^{(l)}_k$ is exactly $v_k(S^{(l)}_k)$. \end{subproof}

\begin{lem_app}{lem_doublinglemma}
(Doubling Lemma) Consider any project $k$ and the set of elements assigned to $k$ in our solution ($S_k$). Let $T$ be some set of agents such that $T \cap S_k  = \emptyset$ and $|T| > |S_k|$. Then, the total payment to the agents in $T \cup S_k$ is at least $v_k(S_k \cup T)$, i.e.,
$$\sum_{i \in T \cup S_k} \bar{p}_i \geq v_k(S_k \cup T).$$
\end{lem_app}
\begin{subproof} We assume for convenience that both $|T|$ and $|S_k|$ are even. The same proof holds when one or both the sets are odd with a few minor modifications. We introduce some additional notation: recall that $T^>$ consists of the elements of $T$ in the decreasing order of their marginal (or payment). We partition $T^>$ into two sets $T_1$ and $T_2$ containing the first $\frac{|T|}{2} + \frac{|S_k|}{2}$  elements of $T^>$, and the last $\frac{|T|}{2} - \frac{|S_k|}{2}$ elements respectively (see Figure~\ref{fig_doublingpartition}). Let $x$ be the agent in $T_2$ who was first assigned to some project during the course of our algorithm. By definition, for every $i \in T_1$, $p_i \geq p_x$, and by the monotonicity of the marginals, for every $i \in T_2$, $p_i \leq p_x$.

\begin{figure}
\centering
\includegraphics[scale=1]{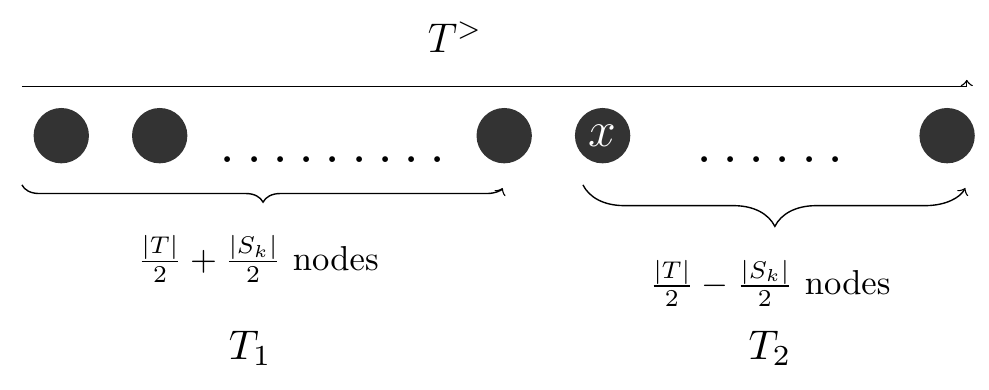}
\caption{The partition of $T^>$ (nodes in $T$ ordered in the decreasing order of the marginals) into two sets $T_1$ and $T_2$}
\label{fig_doublingpartition}
\end{figure}

The rest of the proof proceeds as follows, we first establish lower bounds on $p_x$ and then use the fact that $p_i \geq p_x$ for all $i \in T_1$ to show that the payments are large enough. Suppose that agent $x$ was assigned to some project during iteration $l$ of our algorithm. Let $Rem$ be the agents in $S_k$ who were unassigned before the $l^{th}$ iteration, i.e., $Rem = S_k \setminus S^{(l)}_k$. Since the greedy algorithm \emph{did not choose} to assign the agents in $T_2 \cup Rem$ to project $k$ during iteration $l$, this means that the marginal contribution of the agents chosen in iteration $l$ by our algorithm ($p^{(l)} = p_x$) is at least the average welfare due to the alternative assignment that was not chosen, i.e.,

$$p_x \geq \frac{v_k(T_2 \cup Rem ~ | ~ S^{(l)}_k)}{|Rem| + |T_2|} = \frac{v_k(S_k \cup T_2) - v_k(S^{(l)}_k)}{|Rem|+ |T_2|}.$$

Since $|Rem| \leq |S_k|$ and $|T_2| = \frac{|T|}{2} - \frac{|S_k|}{2}$, we get that $|Rem|+ |T_2| \leq \frac{1}{2}(|T| + |S_k|)$. Substituting this in the above inequality, and bringing the denominator over to the other side, we obtain

$$(|T|+|S_k|)p_x \geq 2(v_k(S_k \cup T_2) - v_k(S^{(l)}_k)).$$ Look at the set $S_k \cup T_2$, the cardinality of this set is exactly $\frac{1}{2}(|T|+|S_k|)$. Therefore, from our fundamental Proposition~\ref{prop_anon_fundamental}, we know that $v_k(S_k \cup T_2) \geq \frac{1}{2} v_k(S_k \cup T)$. Therefore, we get the following final lower bound on $p_x$ multiplied by $|T|+|S_k|$,

$$(|T| + |S_k|)p_x \geq v_k(S_k \cup T) - 2v_k(S^{(l)}_k) \geq v_k(S_k \cup T) - 2v_k(S_k).$$

The last inequality comes from the fact that $S^{(l)}_k \subseteq S_k$. Remember that the final payment to every agent is twice her marginal. Moreover, for every agent in $T_1$, her initial payment is at least $p_x$. Now, we can finally prove the desired lemma,
\begin{align*}
\sum_{i \in S_k \cup T}\bar{p}_i &\geq 2\sum_{i \in S_k}p_i + 2\sum_{i \in T_1}p_i \\
& \geq 2v_k(S_k) + 2|T_1|p_x \\
& = 2v_k(S_k) + (|T|+|S_k|)p_x \\
& \geq 2v_k(S_k) + v_k(S_k \cup T) - 2v_k(S_k)\\
& = v_k(S_k \cup T).
\end{align*}
\end{subproof}

%

\subsubsection*{Fair Payments}
Given a payment vector $\vec{p}$ and an allocation $S$, we say that the payments are fair or envy-free, if for every $i,j \in \mathcal{N}$ such that both $i$ and $j$ belong to project $k$ in $S$, then $p_i=p_j$.

\begin{clm_app}{clm_envyfreeanonymous}
For any instance where the projects have anonymous subadditive valuations, there exists a $(2,2)$-core stable solution such that the payments are envy-free, i.e., all the agents assigned to a single project receive the same payment.
\end{clm_app}
\begin{mainproof}
Once again, we run Algorithm~\ref{alg_greedyanon} to obtain an allocation $S$, but the payments that we provide are defined as follows: for every agent $i \in S_k$, her payment is now $p_i^{(ef)} = 2\frac{v_k(S_k)}{|S_k|}$. From the proof of Theorem~\ref{thm_anonsubadditive}, the solution is still a $2$-approximation to $OPT$. Moreover, the payments are clearly $2$-budget balanced. It only remains for us to show that no group of agents $T$ can deviate to any project $k$ and collectively improve their utility. Instead of proving this from scratch, we piggyback on the proof of Theorem~\ref{thm_anonsubadditive} and reduce our stability condition to that of the Theorem. More specifically, suppose that $(\bar{p})_{i \in \mathcal{N}}$ denotes the payment vector used in Theorem~\ref{thm_anonsubadditive}. Then we show that as long as the solution $(S, (\bar{p})_i)$ is a $2$-core, so is the envy-free solution $(S, \vec{p^{(ef)}})$. The following lemma is the basic building block that facilitates this reduction.

\begin{lemma}
\label{lem_comparitivepaymentsanon}
For a given project $k$, and any positive integer $t \leq |S_k|$, let $T$ denote the set of $t$ agents in $S_k$ with the smallest payments according to $(\bar{p})_{i \in \mathcal{N}}$, i.e., for any $i \in T$, and any $j \in S_k \setminus T$, $\bar{p}_j \geq \bar{p}_i$. Then,
$$\sum_{i \in T}\bar{p}_i \leq \sum_{i \in T}p_i^{(ef)} = 2t\frac{v_k(S_k)}{|S_k|}.$$
\end{lemma}
\begin{subproof}
Assume by contradiction that the above inequality is false, i.e., $\sum_{i \in T}\bar{p}_i > 2t\frac{v_k(S_k)}{|S_k|}$. Then, there must exist at least one agent $i \in T$ whose payment $\bar{p}_i$ is strictly larger than $2\frac{v_k(S_k)}{|S_k|}$. This implies that for every agent $j$ in $S_k \setminus T$, her payment is also strictly larger than the above quantity. Therefore, we have
$$\sum_{i \in S_k}\bar{p}_i = \sum_{i \in T}\bar{p}_i + \sum_{i \notin T}\bar{p}_i > 2t\frac{v_k(S_k)}{|S_k|} + 2(|S_k|-t)\frac{v_k(S_k)}{|S_k|}.$$

However, this is a contradiction since we know that $\sum_{i \in S_k}\bar{p}_i = 2v_k(S_k)$. \end{subproof}

Now, we are ready to show that for every project $k$, and every group $T$, $\sum_{i \in T\cup S_k}p_i^{(ef)} \geq v_k(T\cup S_k)$. Without loss of generality, we can assume that from every project $l$ only the members with the lowest payments under the solution $(\bar{p})_{i \in \mathcal{N}}$ deviate. That is, suppose that $T$ consists of some $t$ members originally in project $l$, then we can assume that these coincide with the $t$ members of $S_l$ who had the lowest payments in $(\bar{p})_{i \in \mathcal{N}}$. Now, from Lemma~\ref{lem_comparitivepaymentsanon}, we know that

$$\sum_{i \in T}p^{(ef)}_i = \sum_{l \in \mathcal{P}}\sum_{i \in T \cap S_l}p^{(ef)}_i \geq \sum_{l \in \mathcal{P}}\sum_{i \in T \cap S_l}\bar{p}_i = \sum_{i \in T}\bar{p}_i.$$

Since $\sum_{i \in S_k}\bar{p}_i = \sum_{i \in S_k}p^{(ef)}_i = 2v_k(S_k)$, this means that $\sum_{i \in T\cup S_k}p_i^{(ef)} \geq \sum_{i \in T\cup S_k}\bar{p}_i$. However, from the proof of Theorem~\ref{thm_anonsubadditive}, we know that $\sum_{i \in T\cup S_k}\bar{p}_i \geq v_k(T\cup S_k)$. This completes the proof.
\end{mainproof}

\section{Appendix: Proofs for Submodular and Fractionally Subadditive Valuations}

\subsection*{XoS Oracles} One of our results for this section requires access to an XoS oracle~\cite{dobzinskiNS10}, which when queried with a XoS valuation $v_k$ and a set $S_k$ returns the XoS clause $a^{(l)}_k$ that maximizes $a^{(l)}_k(S_k)$. In addition, we also assume the presence of a Demand Oracle for each function. For any XoS function, both of these oracles can be implemented in time polynomial in the size of the input (number agents, number of independent clauses).

\begin{prop_app}{prop:XoSexistence}
There exists a $(1,1)$-core stable solution for every instance where the projects have XoS valuations.
\end{prop_app}
Since Submodular functions are contained in the XoS class, this result extends to Submodular valuations as well.
\begin{mainproof}
The proof is simple: let $O=(O_1, \ldots, O_m)$ be the welfare maximizing solution. We define the following procedure to award payments to each agent: for each project $k$, invoke the XoS oracle for $v_k$, $O_k$ and suppose that the returned additive clause is $a^{(l)}_k$. For every $i \in S_k$, her payment is exactly $\bar{p}_i = a^{(l)}_k$. Notice that by definition of the XoS class, for every project $k$, and set $T \subseteq S_k$, $a^{(l)}_k(T) = \sum_{i \in T}\bar{p}_i \leq v_k(T)$.

We now claim that these payments along with the allocation $O$ form a core stable solution. Clearly the solution is budget-balanced, so we only need to show that no group of agents can deviate. Assume to the contrary, and suppose that some set $T$ of agents can deviate to project $k$ and collectively improve their utility. Let $O'$ be the allocation obtained from $O$ by the deviation of the agents in $T$ to project $k$. Then, we show that the social welfare of $O'$ must be strictly larger than the social welfare of $O$, which is a contradiction, i.e.,
\begin{align*}
SW(O') & = v_k(S_k \cup T) + \sum_{l \neq k}v_l(O'_l)\\
& > \sum_{i \in S_k \cup T}\bar{p}_i + \sum_{l \neq k}\sum_{i \in S_l \setminus T}\bar{p}_i
= \sum_{i \in \mathcal{N}}\bar{p}_i = SW(O^*).
\end{align*}
\end{mainproof}

\begin{thm_app}{thm_mainxossubmod}
\begin{enumerate}
\item For any instance where the projects have XoS valuations, we can compute $(1+\epsilon,\frac{e}{e-1})$-core stable solution using Demand and XoS oracles, and a $(\frac{e}{e-1},\frac{e}{e-1})$-core stable solution without these oracles.

\item For submodular valuations, we can compute a $(1+\epsilon,\frac{e}{e-1})$-core stable solution using only a Value oracle.
\end{enumerate}
\end{thm_app}

\begin{mainproof}
\subsection*{(Part 1.1) Xos Valuations with Demand, Xos Oracles}
We begin with some additional notation. Given an allocation $X$, we define the `XoS price' of every agent $i$, $p_i(X)$ as follows: suppose that $i \in X_k$, then use the XoS oracle and obtain the maximizing additive clause $a^{(l)}_k$ for  $X_k$. Set $p_i(X) = a^{(l)}_k(i)$, i.e., the value of agent $i$ in the additive valuation that obtains the maximum value for the allocation $X_k$. It is not hard to see that for every project, the sum of XoS prices of the agents assigned to that project is exactly the welfare due to that project, i.e., $\forall k \in \mathcal{P}$,
$$\sum_{i \in X_k}p_i(X) = a^{(l)}_k(X_k) = v_k(X_k).$$

The algorithm that we present is quite intuitive and based on a best response approach using the social welfare as a potential function.
\begin{enumerate}
\item Initialize the input allocation $A$ to be the $\frac{e}{e-1} ~\approx 1.58$ approximation to OPT obtained using the Algorithm of~\cite{dobzinskiNS10}.

\item Let $p_i(A)$ denote the XoS price of every agent $i$ for the allocation $A$.

\item Define the current allocation $X = A$, and current payments $p_i = p_i(A) + \frac{\epsilon}{N}SW(A)$.

\item Perform a best-response step, i.e., allow a set $T$ of agents to deviate to project $k$ if we can simultaneously improve all of their payments, i.e., if $v_k(T | X_k) > \sum_{i \in T}p_i$.

\item Update the current allocation $X$, and the current payments to be $p_i = p_i(X) + \frac{\epsilon}{N}SW(X)$.

\item When no such deviation is possible, return the allocation $S=X$, and the final payments $(\bar{p})_{i \in \mathcal{N}} = \vec{p}$.
\end{enumerate}

We need to show the following: $(i)$ for a given $\epsilon > 0$, the algorithm converges after making a polynomial number of queries to the Demand and XoS oracles, $(ii)$ the solution returned has at least the welfare of the initial allocation $A$, $(iii)$ the solution is $(1+\epsilon)$-Core stable.

We begin with an easy observation: after every iteration of our algorithm, the total payment made to all the agents is at most a factor $(1+\epsilon)$ times the current social welfare. This is because the payment to every agent is her XoS price plus $\frac{\epsilon}{N}$ times the current welfare; since there are a total of $N$ agents we get that the total payments is the aggregate XoS price plus $\epsilon$ times the social welfare. Our next claim is that in every iteration, our algorithm makes at most $m$ calls to the XoS oracle and to the Demand oracle. Observe that we require the XoS oracle only when we choose payments for the agents (invoke the XoS prices), the demand oracle is only required for identifying a best-response step in our algorithm (Step (5)). In order to find the XoS prices, it suffices if we invoke the XoS oracle once for each every project $k$, given the allocation $X_k$. Next, in order to check for a best-response step, we need to see if $\exists$ $k, T$ such that $v_k(T | X_k) > \sum_{i \in T}p_i$. This can be done using the demand oracle for project $k$ using the price vector $\vec{p}$, and the reduced valuation function $v_k(.|X_k)$. Therefore, in order to determine whether there exists a best-response step, we need to call the Demand oracle at most $m$ times, once for each project. This completes the proof of the claim.

Now that we know that our algorithm makes at most $m$ calls to each of the oracles in each iteration, if we can prove that our algorithm converges after a polynomial number of iterations, then the total number of queries to the oracles is also polynomially bounded. In order to show this, we first prove a more fundamental lemma that gives a lower bound on the increase in social welfare during each iteration.
\begin{lemma}
\label{lem_xos_welfareincr}
In every iteration, the social welfare of the current allocation increases strictly by at least a fraction $\frac{\epsilon}{N}$ of current social welfare.
\end{lemma}
That is, suppose that during some iteration, the initial allocation was $X^1$, and the allocation after the best-response was $X^2$, then $SW(X^2) > SW(X^1)(1+\frac{\epsilon}{N})$.

\begin{subproof}
Suppose that during this allocation, some set $T$ of agents deviate to project $k$. Let the initial payments of all the agents (before the deviation) be $\vec{p}$. We also know by definition of the best-response step that $v_k(T | X^1_k) > \sum_{i \in T}p_i$. Moreover, it is not hard to see that (initially) for every project $l$, the payments are at most the welfare due to the project plus a markup, i.e., $\sum_{i \in X^1_l}p_i = v_l(X^1_l) + |X^1_l|\frac{\epsilon}{N}SW(X^1).$

Our next observation establishes a relationship between the welfare of any project $l$ after the best-response step, and the initial set of prices. We know that for all $l \in \mathcal{P}$ with $l\neq k$, $X^2_l \subseteq X^1_l$. Then, we show that:

$$v_l(X^2_l) \geq \sum_{i \in X^2_l}(p_i - \frac{\epsilon}{N}SW(X^1)).$$

To see why the above inequality is true, suppose that for a given project $l\neq k$, $r_1$ and $r_2$ are the maximizing clauses for the assignments $X^1_l$ and $X^2_l$ respectively. The above inequality follows from the fact that $v_l(X^2_l) = a^{r_2}_l(X^2_l) \geq a^{r_1}_l(X^2_l)$. The final term is exactly the initial XoS prices of the agents in $X^2_l$, which is $p_i - \frac{\epsilon}{N}SW(X^1)$ for every agent $i$. Now we are in a position to show our desired result,
\begin{align*}
SW(X^2) &=  v_k(X^2_k) + \sum_{l \neq k}v_l(X^2_l) \\
& \geq \left[ v_k(X^1_k) + v_k(T | X^1_k)\right] + \sum_{l \neq k}\sum_{i \in X^2_l}(p_i - \frac{\epsilon}{N} SW(X^1))\\
& > v_k(X^1_k) + \sum_{i \in T}p_i + \sum_{l \neq k}\sum_{i \in X^2_l}(p_i - \frac{\epsilon}{N} SW(X^1)).
\end{align*}

Now we can rearrange the above inequality by assigning every agent $i \in T$ to the project that $i$ belonged to in $X^1$, and then rewriting $p_i$ for these agents as $p_i - \frac{\epsilon}{N}SW(X^1) + \frac{\epsilon}{N}SW(X^1)$. This gives us,
\begin{align*}
SW(X^2) & > v_k(X^1_k) + \sum_{l \neq k}\sum_{i \in X^1_l}(p_i - \frac{\epsilon}{N} SW(X^1)) + \frac{\epsilon}{N}|T|SW(X^1)\\
& \geq v_k(X^1_k) + \sum_{l \neq k}v_l(X^1_l) + \frac{\epsilon}{N}SW(X^1)\\
& = SW(X^1) + \frac{\epsilon}{N}SW(X^1).
\end{align*}
This completes the proof of the lemma.
\end{subproof}

Now, we can use the above lemma to prove the main theorem. We begin by showing that our algorithm converges after a polynomial number of iterations. Clearly, the social welfare of the current allocation is non-decreasing throughout our algorithm, and therefore bounded from below by $SW(A)$. This means that in every iteration, the increase in welfare is strictly larger than $\frac{\epsilon}{N}SW(A)$. But we also know that $\frac{e}{e-1}SW(A) \geq SW(OPT)$. Therefore, we can bound the number of iterations of our algorithm by the total possible increase in welfare divided by the increase in welfare in each iteration,
$$\frac{SW(OPT) - SW(A)}{\frac{\epsilon}{N}SW(A)} \leq \frac{N}{(e-1)\epsilon}.$$

Therefore, our algorithm converges after a polynomial number of calls to the Demand and XoS Oracles. The final welfare, as per Lemma~\ref{lem_xos_welfareincr} is at least the initial welfare, i.e., $SW(A)$. Core stability follows by definition of the best-response phase because if there existed a set $T$, and a project $k$, such that $v_k(T | S_k) > \sum_{i \in T}\bar{p}_i$, then our algorithm would not have terminated. Together with the fact that $\sum_{i\in S_k}\bar{p}_i \geq v_k(S_k)$, we get that $\sum_{i\in T\cup S_k}\bar{p}_i \geq v_k(T | S_k) +v_k(S_k) = v_k(T\cup S_k)$, as desired. Finally, the payments are $(1+\epsilon)$-budget balanced:

$$\sum_{i \in \mathcal{N}}\bar{p}_i = \sum_{k \in \mathcal{P}}(v_k(S_k) + \frac{\epsilon |S_k|}{N}SW(S)) = SW(S) + \epsilon SW(S) \blacksquare$$

\subsection*{(Part 1.2) XoS Valuations without the Oracles}
We actually show a much stronger result here, namely that given a $\alpha$-approximation to the LP optimum, we can obtain a $(\alpha, \alpha)$-Core stable solution. Plugging in $\alpha=\frac{e}{e-1}$ for fractionally subadditive functions, we get the desired result. Note that a $(\alpha,\alpha)$-core result is strictly better than our $(2\alpha,2\alpha)$ result for general subadditive functions.

We only sketch the key features of this proof since the approach is very similar to our algorithm for general subadditive functions as described in Theorem~\ref{thm_mainsubadditive}. Suppose that the input allocation is $A$, and the optimum dual prices are $\vec{p^*}$, $\vec{z^*}$ as usual. Given any allocation $X$, define $marg_i(X)$ to be agent $i$'s marginal contribution to her project, i.e., suppose that agent $i \in X_k$, then define $marg_i(X) = v_k(X_k) - v_k(X_k - \left\lbrace i \right\rbrace)$. Now, our algorithm for this theorem involves the following adaptation of the Greedy Matching procedure described in the proof of Theorem~\ref{thm_mainsubadditive}. Recall that for any allocation $X$, $\zeta(X)$ gives the set of empty projects under that allocation.

\begin{enumerate}
\item Let the current allocation $X$ be initialized to $A$.

\item Allow a single agent $i$ to deviate to some empty project $k \in \zeta(X)$ as long as this leads to a strict increase in social welfare, i.e, $v_k(i) > marg_i(X)$.

\item Update the current allocation $X$ and repeat Step (2) until no agent wants to deviate.

\item Output $S$ to be the (final) current allocation.
\end{enumerate}

Clearly, it is not hard to see that the welfare of the final allocation is not smaller than the welfare of the initial allocation $A$. Moreover, the above algorithm must converge in time Poly$(N,m)$. The crux of the proof lies in the following slightly intricate payment scheme,

\begin{itemize}
\item Define for every project $k$, high value users $(S^H_k)$ and low value users $(S^L_k)$ as follows,
$$S^H_k = \left\lbrace i \in S_k | marg_i(S) > p^*_i \right\rbrace, \quad\quad S^L_k = S_k \setminus S^H_k.$$

\item Define for every project $k$, the residual slack $z_k$ as follows,
$$z_k = \sum_{i \in S^H_k}p^*_i + z^*_k - \sum_{i \in S^H_k}marg_i(S).$$

\item For every project $k$, and every low value agent $i \in S_k$, let her final payment be $\bar{p}_i = p^*_i$.

\item For every project $k$, and every high value agent $i \in S_k$, let her final payment be $\bar{p}_i = marg_i(S) + \frac{z_k}{|S^H_k|}.$
\end{itemize}

The following is the crucial lemma that helps us show core stability.
\begin{lemma}
\label{lem_xosnooracle}
For every agent $i$, her payment is at least her marginal value. Moreover, the total payment to the members of any project $k$ is exactly $\sum_{i \in S_k}p^*_i + z^*_k$.
\end{lemma}
\begin{subproof}
Let us begin with the first part of the lemma, this is clearly true for low value agents by definition. Therefore, we only need to focus on the high value agents. Since every high value agent's payment is her marginal value plus some fraction of the residual slack, it is enough if we show that for every project $k$, the residual slack is positive. We use the following property of XoS functions: given assignment $S_k$, let $l$ be the clause that maximizes $a^{(l)}_k(S_k)$, then, $marg_i(S) = v_k(S_k) - v_k(S_k - \left\lbrace i \right\rbrace) \leq a^{(l)}_k(S_k) - a^{(l)}_k(S_k - \left\lbrace i \right\rbrace) = a^{(l)}_k(i).$ Now we can prove the first part of the lemma,
\begin{align*}
z_k & = \sum_{i \in S^H_k}p^*_i + z^*_k - \sum_{i \in S^H_k}marg_i(S)\\
& \geq v_k(S^H_k) - \sum_{i \in S^H_k}a^{(l)}_k(i) \\
& \geq v_k(S^H_k) - v_k(S^H_k) = 0.
\end{align*}
This completes the first part of the lemma. The second part follows from definition of the payments.
\end{subproof}

Now the rest of the proof follows in the same fashion as that of Theorem~\ref{thm_mainsubadditive}. Clearly, the final payments are exactly equal to the value of the Dual Optimum which is a factor $\alpha$ larger than the welfare of the current solution. In order to show that no group of agents $T$ can deviate to any project for subadditive valuations (remember that XoS is a sub-class of subadditive valuations), it is enough to show the following (See proof of Theorem~\ref{thm_mainsubadditive}) $(i)$ every agent $i$'s payment is at least $p^*_i$, $(ii)$ for every project $k$, the total payment of the members of that project is at least $\sum_{i \in S_k}p^*_i + z^*_k$, $(iii)$ if $k$ is an empty project, then for any agent $i$, $\bar{p}_i \geq v_k(i)$. $(i)$ and $(ii)$ guarantee that any deviation to a non-empty project $k$ will not occur due to dual feasibility, and $(iii)$ together with subadditivity guarantees that no set benefits from deviating to an empty project. The first two requirements follow immediately from the definition of the payments and the above Lemma. For the third requirement, observe that at the end of the payment algorithm defined above, $marg_i(S) \geq v_k(i)$ for $k \in \zeta(S)$. Therefore, as per Lemma~\ref{lem_xosnooracle}, $\bar{p}_i \geq marg_i(S) \geq v_k(i)$. This completes the proof.

\subsection*{(Part 2) Submodular Functions with only Value Queries}

The algorithm is exactly the same as in the proof for XoS functions in Part 1.1; recall that submodular functions are a strict sub-class of XoS functions. The key difference here is that we can eliminate the dependence on both Demand and XoS oracles using the nice properties of Submodular functions. First, note that we use the XoS oracle to find the corresponding additive clause for a given valuation function $v_k(S_k)$. It is known that one can implement XoS queries in polynomial time for Submodular functions using a simple Greedy Approach as in Claim~\ref{clm_submodular_warmup}.

More specifically, given a submodular function $v_k$, and a corresponding set $S_k$, the goal of an XoS oracle query is to obtain an additive function $a$ such that $v_k(S_k)=a(S_k)$ and for all $T\subseteq S_k$, $v_k(T)\geq a(T)$. It is not hard to see that this gives us the XoS clause which is maximum at $S_k$ without loss of generality. For submodular functions, we can compute the XoS prices as follows: $(i)$ order the elements of $S_k$ in some arbitrary order, $(ii)$ add the elements of $S_k$ to $v_k(\emptyset)$ one after the other in the predetermined order, $(iii)$ agent $i$'s XoS price is the marginal cost of adding her to the set, i.e., if $A_i$ is the set of elements before $i$ in the predetermined order, then agent $i$'s XoS price is $v_k(A_i \cup \left\lbrace i \right\rbrace) - v_k(A_i)$. These prices give us the additive function $a$, as desired.

Now, we move on to Demand Oracles whose main purpose is to identify in every iteration, a project $k$, and a set $T$ such that $v_k(T|X_k) > \sum_{i \in T} p_i$, where $X_k$ is the set of agents currently assigned to that project. For submodular functions, instead of looking at group deviations to a project, it suffices if we look at individual deviations. More specifically, our claim is the following: if $\exists T$ satisfying $v_k(T|X_k) > \sum_{i \in T} p_i$, then there exists at least one agent $i \in T$ satisfying $v_k(X_k \cup \left\lbrace i \right\rbrace ) - v_k(X_k) > p_i$. This follows almost directly from submodularity,

$$\sum_{i \in T}p_i < v_k(T|X_k) < \sum_{i \in T}v_k(X_k \cup \left\lbrace i \right\rbrace) - v_k(X_k).$$

Therefore, our claim must be true for at least one such agent. The claim has the following implication: the best-response algorithm for Submodular functions can be obtained from the B-R algorithm for XoS valuations in Part 1.1 by changing step (4) to ``allow a single $i$ to deviate to project $k$ as long as $v_k(X_k \cup \left\lbrace i \right\rbrace) - v_k(X_k) > p_i$". Whether or not such a deviation exists can be found using $O(Nm)$ value queries. Finally, we remark that a $\frac{e}{e-1}$-approximation to $OPT$ that we use as an input to the algorithm can be computed for submodular functions using only Value Queries~\cite{vondrak08}.

The rest of the proof follows.
\end{mainproof}


\section{Appendix: Proofs from Section 4}

\begin{thm_app}{thm_equivalencecombauc}
Every Core stable solution for a given instance of our utility sharing game can be transformed into a Pure Nash Equilibrium (with weak no-overbidding) of the corresponding `flipped' simultaneous second price auction, and vice-versa.
\end{thm_app}
\begin{mainproof}
Suppose that we are given an instance $(\mathcal{N}, \mathcal{P}, (v)_{k \in \mathcal{P}})$ of our combinatorial utility sharing game along with a core stable solution $(S, (\bar{p})_{i \in \mathcal{N}})$. We construct a bid profile $\vec{b} = (\vec{b_1}, \ldots, \vec{b_m})$ for the flipped combinatorial auction that in combination with the allocation $S$, and the corresponding second-price payments constitutes a Pure Nash equilibrium with no weak-overbidding. Our construction is simple, for every buyer $k \in \mathcal{P}$, and item $i \in \mathcal{N}$: $b_k(i) = \bar{p}_i$ if $i \in S_k$ and $b_k(i) = 0$ otherwise. Notice that for every item, only one buyer (the winning buyer) has a positive bid, therefore,
the corresponding auction mechanism will output $S$ as the winning allocation along with zero payments. Therefore, every every buyer $k$'s utility is exactly $v_k(S_k)$.

In order to show that this is a pure Nash equilibrium, it is enough to prove for a fixed buyer $k$, and for any alternative bid vector $\vec{b'_k}$, buyer $k$'s utility cannot strictly improve at her new bid when all the other buyers bid according to the constructed profile. Suppose that at the new bid vector, player $k$ wins the set $S'_k$ of items and define $T = S'_k \setminus S_k$. Then, for every item $i \in T$, buyer $k$ has to pay exactly $\bar{p}_i$ (the bid of the second highest bidder) in the new auction solution. Therefore, player $k$'s new utility is
$$u_k(\vec{b'_k}, \vec{b_{-k}}) = v_k(S'_k) - \sum_{i \in T}\bar{p}_i \leq v_k(S_k \cup T) - \sum_{i \in T}\bar{p}_i.$$

Now upon applying the core stability requirement for the empty set, we get $v_k(S_k) \leq \sum_{i \in S_k}\bar{p}_i$. This condition must be true for every project $l \in \mathcal{P}$. Moreover, this condition in combination with budget-balance indicates that for project $k$ (and every other project), $v_k(S_k) = \sum_{i \in S_k}\bar{p}_i$. Now, we can bound the player's new utility using core-stability as

$$v_k(S_k \cup T) - \sum_{i \in T}\bar{p}_i \leq \sum_{i \in S_k}\bar{p}_i = v_k(S_k).$$

The last quantity is player $k$'s original utility, and therefore this deviation cannot strictly benefit the player. Notice that this also implies the no weak-overbidding condition.

Now, for the other direction, suppose that we are given a winning allocation $S$ along with a bid profile $\vec{b}$ that together forms a Nash Equilibrium (the second-price payments $\vec{P}$ are implicit in the bid). Construct the following payment vector: for every item $i \in \mathcal{N}$, $p_i = \max_{k \in \mathcal{P}} b_k(i)$, i.e., the highest bid for that item. Consider any player $k$, and a set of items $T$ outside of her current allocation. In order to win these items as well as the items she received in $S$, player $k$ must bid at least $p_i + \epsilon$ for every $i \in T$ without changing her bids for the items in $S_k$. Let $\vec{b'_k}$ be the modified bid vector reflecting these increased bids for the items in $T$. However, since this is a Nash equilibrium, the utility of the player under the modified bid vector cannot be larger than her original utility, i.e.,
\begin{align*}
u_k(\vec{b'_k},\vec{b_{-k} }) & = v_k(S_k \cup T) - P(S_k) - \sum_{i \in T}p_i \\
& \leq u_k(\vec{b})\\
& = v_k(S_k) - P(S_k).
\end{align*}
Therefore, we have for every buyer $k$, and every set of items $T$, $v_k(S_k \cup T) - \sum_{i \in T}p_i \leq v_k(S_k)$. Due to the weak no-overbidding assumption, we know that for each $k$, $\sum_{i\in S_k}p_i \leq v_k(S_k)$. We now set the final payments $\bar{p}_i$ by increasing the payments $p_i$ arbitrarily for each $k$ until $\sum_{i\in S_k}\bar{p}_i = v_k(S_k)$. These payments are clearly budget-balanced. Moreover, since we only increased the payments, we still have that for every buyer $k$, and every set of items $T$, $v_k(S_k \cup T) - v_k(S_k) \leq \sum_{i \in T}\bar{p}_i$. From this inequality, it is easy to see that the solution $S$ along with the payments $(\bar{p})_{i \in \mathcal{N}}$ constitutes a core-stable solution. This solution retains the welfare of the original Nash equilibrium.
\end{mainproof}

\subsection*{Instance where every Nash Equilibrium requires a large amount of overbidding}
The example that we use here is the same as the instance for Part (2) of Claim~\ref{clm_badexamplesubadd}. Consider an auction with just two buyers having the following anonymous subadditive valuations: $v_1(S) = \frac{N}{2} ~ \text{for }S \neq \mathcal{N}$, $v_1(\mathcal{N}) = N$; and $v_2(S) =  \sqrt{N} ~~ \forall S \subseteq \mathcal{N}.$ We show that in any Pure Nash Equilibrium $(\vec{b_1},\vec{b_2})$ and $S$, at least one buyer has to over-bid by a factor $O(\sqrt{N})$. First, consider any Nash equilibrium where buyer $1$ wins all of the items. Then, it is not hard to see that $b_1(i) \geq \sqrt{N}$ for every $i \in \mathcal{N}$ or else buyer $2$ can bid for $i$ $(b'_2(i)=\sqrt{N})$, win the item and strictly improve her utility. Therefore, we have $b_1(\mathcal{N}) \geq \sqrt{N} \times N = \sqrt{N} v_1(\mathcal{N})$, which gives us the amount by which buyer $1$ overbids.

Next, suppose that in a Nash Equilibrium of this auction, buyer $2$ receives some ($S_2$) but not all the items. Then, buyer $1$'s utility is $v_1(S_1) - b_2(S_1) = \frac{N}{2} - b_2(S_1)$. Now what if buyer $1$ modifies her bids in order to win all the items in $\mathcal{N}$ and pays $b_2(\mathcal{N})$ for the same? Such a deviation cannot improve her utility, i.e., $N - b_2(\mathcal{N}) \leq \frac{N}{2} - b_2(S_1)$. Simplifying this, we get $b_2(S_2) \geq \frac{N}{2}$. Since $v_2(S_2) = \sqrt{N}$, we get that the amount of over-bidding by buyer $2$ is $\frac{N}{2\sqrt{N}} = \frac{\sqrt{N}}{2}$. It is not hard to see that this must be the case even when buyer $2$ wins all of the items in $\mathcal{N}$.\\

\begin{clm_app}{clm_combauc_comp}
Given a Second Price Simultaneous Combinatorial Auction with Item Bidding, we can compute
\begin{enumerate}
\item A $2$-approximate Nash Equilibrium that extracts half the optimal social welfare as long as the buyers have anonymous subadditive valuations.
\item A $1+\epsilon$-approximate Nash equilibrium that is a $\frac{e}{e-1}$-approximation to the optimum welfare when the buyers have submodular valuations.

\end{enumerate}

The first solution involves $4$-conservative bids, and the second solution involves $(1+\epsilon)$-conservative bids.
\end{clm_app}

\begin{mainproof}
(Proof of Statement 1) Once again, we turn to our greedy algorithm (Algorithm~\ref{alg_greedyanon}). Consider the solution $S$ returned by our greedy algorithm and envy-free payments $(p^{(ef)})_{i \in \mathcal{N}}$ as per Claim~\ref{clm_envyfreeanonymous}. Convert the payments into a bid profile $\vec{b}$ using the transformation mechanism described in the proof of Theorem~\ref{thm_equivalencecombauc}. Clearly, the solution has the some social welfare as the original allocation, so we only need to prove stability and quantify the level of over-bidding. As with the proof of Theorem~\ref{thm_equivalencecombauc}, suppose that some buyer $k$ changes his bids so that his new winning set is $S'_k$ and $T = S'_k \setminus S_k$. Then we have that the player's new utility is at most

$$v_k(S_k \cup T) - \sum_{i \in T}\bar{p}_i \leq \sum_{i \in S_k}\bar{p}_i = 2v_k(S_k)$$

Therefore, the player's new utility is at most twice her original utility, and therefore, the solution is a $2$-approximate Nash Equilibrium. Now, it only remains for us to prove that the bid vector is $4$-conservative for every buyer $k$.

Consider some $k \in \mathcal{P}$, it suffices to show that $\sum_{i \in T}b_k(i) \leq 4v_k(T)$ holds only for the case when $T$ is subset of $S_k$. This is because the player's bid is zero for every item outside of her winning set. Suppose that $|T|=t$, and let $r$ be a non-negative integer satisfying, $2^rt \leq |S_k| \leq 2^{r+1}t$. We know that player $k$'s bid for every item in $S_k$ is exactly $\frac{2v_k(S_K)}{|S_k|}$. Abusing notation, we denote by $v_k(2^r t)$ the value of any set of size $2^r t$. Now, from a repeated application of the Fundamental Proposition for Anonymous Functions (Proposition~\ref{prop_anon_fundamental}), we get
$$v_k(T) \geq \frac{v_k(2^r t)}{2^r}.$$

Now we are ready for the final leg of our proof.
\begin{align*}
\sum_{i \in T}b_k(i) & = \frac{2v_k(S_k)}{|S_k|} t \\
& \leq 2\frac{v_k(S_k) t}{2^{r}t}  = 2\frac{v_k(S_k)}{2^{r}}\\
& \leq 2\frac{2v_k(2^r t)}{2^{r}} & \text{(From Proposition~\ref{prop_anon_fundamental})}\\
& \leq 4\frac{v_k(T)2^r}{2^{r}} = 4v_k(T)
\end{align*}

(Proof of Statement 2) One could be tempted to claim that the Algorithm in Theorem~\ref{thm_mainxossubmod} yields a $1+\epsilon$-approximate equilibrium (even for XoS valuations), and indeed it does. Unfortunately, in the final payments, every agent is provided her XoS clause price along with an additive mark-up of $\epsilon SW(S)$. Therefore, the level of multiplicative over-bidding (as measured by $\gamma$-conservativeness) for this solution could be quite large. However, we only need to make a few high-level modifications to the Algorithm for Submodular valuations in Theorem~\ref{thm_mainxossubmod} that will allow us to improve the conservativeness while retaining the approximation factor.

The key step in this modified algorithm is how we price items (using the XoS prices) in Step (5) of the algorithm. In particular, the property that we require from the pricing mechanism is the following: after any best-response move that leads to a strict improvement of the social welfare, the price of any given item cannot decrease. A simple pricing mechanism for Submodular functions that achieves this property was described in~\cite{dobzinskiFK15}, the reader is asked for refer to Claim 2.5 of that paper for the exact construction. At a high level, the pricing mechanism works as follows: in every iteration, every buyer maintains an ordering on the item set, which is dynamically updated. Suppose that for a given item $i$ assigned to buyer $k$ at some instance, $A_i$ denotes the set of items before $i$ in buyer $k$'s ordering. Then, the price of item $i$ is exactly $v_k(\left\lbrace i \right\rbrace | A_i)$.

We simply use the above mechanism as a black-box here. Now, we are in a position to describe our modified algorithm: in every round we maintain a single price (or payment) for each item. At the end, we derive bids for the players using the same idea as in Theorem~\ref{thm_equivalencecombauc}.

\begin{enumerate}
\item Initialize the input allocation $A$ to be the $\frac{e}{e-1} ~\approx 1.58$ approximation to OPT obtained using the Algorithm of~\cite{vondrak08}.

\item Let $p_i(A)$ denote the Submodular price of every item $i$ for the allocation $A$ obtained using the mechanism of~\cite{dobzinskiFK15}.

\item Define the current allocation $X = A$, and current payment $p_i = p_i(A)(1+\epsilon)$.

\item Perform a best-response step, i.e., allow an item $i$ to deviate to project $k$ as long as $v_k( i | X_k) > p_i$.

\item Update the current allocation $X$, and let $p_i(X)$ be the current submodular prices obtained using the mechanism of~\cite{dobzinskiFK15}. Then, every agent $i$'s payment is $p_i = p_i(X)(1+\epsilon)$.

\item When no such deviation is possible, return the allocation $S=X$, and the final payments $(\bar{p})_{i \in \mathcal{N}} = \vec{p}$.
\end{enumerate}

We first argue for core-stability before showing the properties of the approximate Nash equilibrium. Clearly, the solution returned is (approximately) core-stable, and has a social welfare that is at least a $\frac{e}{e-1}$-approx to OPT (since the best-response is welfare increasing). We now show bounds on the running time. Notice that in every round, the price of at least one item is increasing by a multiplicative factor of $1+\epsilon$, i.e., $i$ deviates to project $k$ because $v_k(i | X_k) > p_i$, then her price becomes $v_k(i | X_k)(1+\epsilon)$.

Given this, it is not hard to see that the algorithm proceeds for at most $O(N log_{(1+\epsilon)} (\frac{v_{max}}{\Delta}))$ since no agent's price can be strictly larger than $v_{max}$ by definition. We need not worry about agents whose initial payment is zero; either their final payment is still zero or their payment (during some iteration) increases up to $\Delta$, and then multplicatively increases in each subsequent round where it deviates. Moreover, since $log_{(1+\epsilon)(}\frac{v_{max}}{\Delta}) = O(\frac{1}{\epsilon}log(\frac{v_{max}}{\Delta}))$, we get the desired run time.

Now, let use the black-box mechanism of Theorem~\ref{thm_mainxossubmod} to convert payments to bids. We first show that the solution is a $(1+\epsilon)$-approximate Nash equilibrium. Observe that for every project $k$, $\sum_{i \in S_k}b_k(i) = \sum_{i \in S_k}\bar{p}_i = (1+\epsilon)v_k(S_k)$. Now suppose that buyer $k$ modifies her bid, receives an additional set $T$ of items, then her new utility is
$$v_k(S_k \cup T) - \sum_{i \in T}\bar{p}_i \leq \sum_{i \in S_k}\bar{p}_i = (1+\epsilon)v_k(S_k),$$ where $v_k(S_k)$ is her old utility. Therefore, the solution is a $(1+\epsilon)$-approximate equilibrium. Next, we prove that the bids are $(1+\epsilon)$-conservative. Recall that (as per the pricing mechanism) for every project $k$, there exists an ordering of agents $(O_k$) so that each agent's (final) price is $(1+\epsilon)$-times her marginal price when adding agents according to that order. In order to show that the bids are conservative, consider any set $T \subseteq S_k$. We argue that

$$\sum_{i \in T}b_k(i) = \sum_{i \in T}\bar{p}_i \leq (1+\epsilon)v_k(T).$$

Indeed, consider the item $i$ in $T$ who appears first in $O_k$. Clearly its payment is at most $v_k(i)$ by submodularity (since it may not be the first item in $O_k$). This argument can be repeated for every agent recursively. Therefore, in the solution that we obtain, the Nash Equilibrium has $(1+\epsilon)$-conservative bids. \end{mainproof}

\end{document}